\documentclass[%
 reprint,
 amsmath,amssymb,
 aps,
]{revtex4-2}

\usepackage{multirow}
\usepackage{graphicx}
\usepackage{dcolumn}
\usepackage{bm}

\begin{document}

\title{Systematic cRPA study of two-dimensional MA$_2$Z$_4$ materials: From unconventional screening to correlation-driven instabilities}

\author{F. Bagherpour$^{1}$}
\author{Y. Yekta$^{1}$}
\author{H. Hadipour$^{1}$}\email{hanifhadipour@gmail.com}
\author{E. \c{S}a\c{s}{\i}o\u{g}lu$^{2}$}\email{ersoy.sasioglu@physik.uni-halle.de}
\author{A. Khademi$^{3}$}
\author{S. A. Jafari$^{4}$}
\author{I. Mertig$^{2}$}
\author{S. Lounis$^{2}$}
\affiliation{$^{1}$Department of Physics, University of Guilan, 41335-1914, Rasht, Iran \\
$^{2}$Institute of Physics, Martin Luther University Halle-Wittenberg, 06120 Halle (Saale), Germany \\
$^{3}$Department of Physics, Sharif University of Technology, Tehran 11155-9161, Iran \\
$^{4}$2nd Institute of Physics C, RWTH Aachen University, 52074 Aachen, Germany}

\date{\today}

\begin{abstract}
Understanding the interplay between screening, electronic correlations, and collective excitations is essential for the design of two-dimensional quantum materials. Here, we present a comprehensive first-principles study of more than 60 MA$_2$Z$_4$ monolayers, encompassing semiconducting, metallic, cold-metallic, magnetic, and topological phases. Using the constrained random phase approximation (cRPA), we compute material-specific effective Coulomb interaction parameters \( U \), \( U' \), and \( J \), including their spatial dependence across distinct correlated subspaces defined by local coordination and crystal symmetry. In semiconducting compounds, long-range nonlocal interactions persist, revealing unconventional screening and suggesting strong excitonic effects beyond simple dielectric models. In cold-metallic systems, sizable long-range Coulomb interactions remain despite the presence of free carriers, highlighting their atypical metallic screening. Among 33-valence-electron compounds, we find \( U_{\mathrm{eff}} > W \) in the $\beta_2$ phase, indicating proximity to charge-density-wave or Mott instabilities. Several V- and Nb-based systems exhibit intermediate-to-strong correlation strength, with \( U/W > 1 \) in multiple cases. Using cRPA-derived Stoner parameters, we identify magnetic instabilities in various V-, Nb-, Cr-, and Mn-based compounds. Finally, selected cold-metallic systems display plasmon dispersions that deviate from the conventional \(\sqrt{q}\) behavior, revealing nearly non-dispersive low-energy modes. These results position MA$_2$Z$_4$ monolayers as a versatile platform for investigating correlation-driven instabilities and emergent collective behavior in two dimensions.
\end{abstract}

\maketitle

Two-dimensional (2D) materials have reshaped the landscape of condensed matter physics and nanotechnology, 
providing an ideal platform to explore quantum phenomena in reduced dimensions~\cite{novoselov2005two,manzeli20172d}. 
Their atomically thin geometry leads to modified dielectric screening, tunable electronic properties, and access
to a rich spectrum of correlation-driven and collective phases. Beyond their applications in electronics and 
optoelectronics, 2D materials have emerged as a promising platform for strongly correlated phenomena such as Mott 
insulating behavior~\cite{wang2020correlated}, quantum spin liquids~\cite{broholm2020quantum}, and flat-band 
superconductivity~\cite{cao2018unconventional}. In many such systems, electron–electron interactions dominate 
over single-particle band effects, making it essential to understand not only the strength but also the spatial 
dependence of Coulomb interactions. These characteristics are strongly influenced by reduced dimensionality, dielectric 
environment, and orbital character, all of which modulate the screening behavior. Consequently, there is growing 
demand for first-principles frameworks capable of quantifying interaction parameters in 2D materials, particularly 
where conventional assumptions about screening no longer hold.

A fundamental factor governing correlation strength in low-dimensional systems is the nature of Coulomb screening. 
In 2D materials, the suppression of dielectric screening enhances its spatial nonlocality relative to bulk 
systems~\cite{cudazzo2011dielectric, andersen2015dielectric,steinke2020coulomb}, resulting in long-range electron–electron interactions 
that significantly influence both quasiparticle excitations and collective modes. These unconventional screening 
characteristics underlie large exciton binding energies in semiconducting monolayers~\cite{ugeda2014giant} and 
flattened plasmon dispersions in cold metallic 2D systems~\cite{da2020universal}. Moreover, reduced screening can
amplify Coulomb repulsion, favoring magnetic order, Mott transitions, or plasmonic instabilities. These considerations 
highlight the need for a material-specific understanding of screening and correlation effects to capture emergent phases 
and guide functional materials design.

A widely used method for determining electronic interactions from first principles is the constrained random 
phase approximation (cRPA)~\cite{aryasetiawan2004frequency,miyake2008ab,csacsiouglu2011effective,van2021random,cunningham2023qs}. 
By decomposing the electronic polarization into low-energy (model) and high-energy (rest) contributions, cRPA 
enables the computation of partially screened Coulomb interactions, such as the on-site intra-orbital Hubbard $U$, 
inter-orbital interaction $U'$, and exchange $J$ that enter effective low-energy Hamiltonians. This approach has
been successfully applied to a broad range of correlated materials, including transition metal oxides~\cite{Zhang_2025}, 
unconventional superconductors~\cite{Yue_2025}, and monolayer TMDs~\cite{Ramezani,acharya2021importance}. However, a systematic investigation 
of cRPA-derived interaction parameters in the recently discovered 2D MA$_2$Z$_4$ materials remains absent, despite 
the structural uniqueness and correlated ground states found in this class.

\begin{figure*}[tp]
\centering
\includegraphics[width=0.97\linewidth]{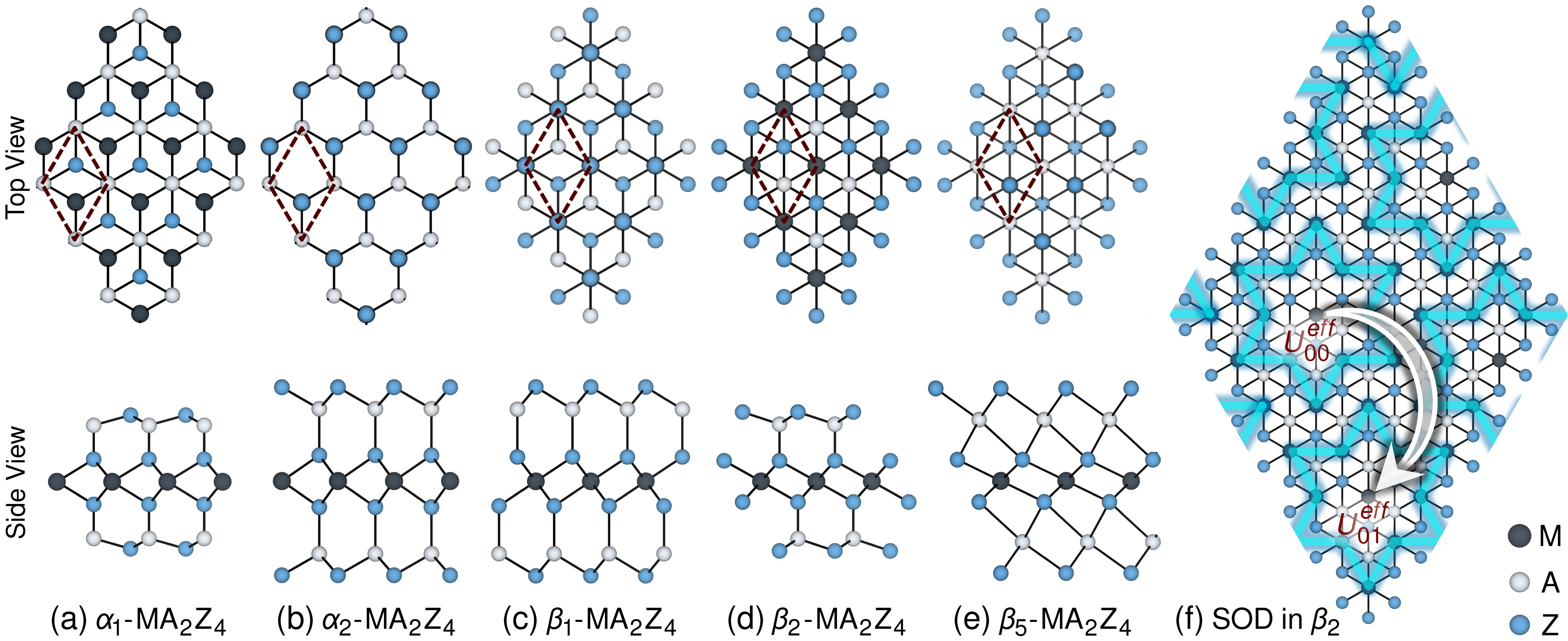}
\caption{Side and top views of the two-dimensional crystal structure of intercalated architecture MA$_2$Z$_4$ in 
(a) $\alpha_1$ structure, (b) $\alpha_2$ structure, (c) $\beta_1$ structure, (d) $\beta_2$ structure, (e) $\beta_5$ 
structure, and (f) Star-of-David (SOD) reconstructed $\beta_2$ crystal structure. Dark gray, white, and blue spheres 
denote M, A, and Z atoms, respectively.}
\label{fig:struc}
\end{figure*}

MA$_2$Z$_4$ monolayers represent a chemically and structurally diverse family of 2D materials that extend 
beyond traditional van der Waals systems~\cite{hong2020chemically, Wang}. Unlike most 2D materials derived 
from layered bulk crystals, MA$_2$Z$_4$ compounds, first realized via the synthesis of MoSi$_2$N$_4$, lack 
a known three-dimensional analogues. Their septuple-layer structure (Z–A–Z–M–Z–A–Z) combines a TMD-like 
MZ$_2$ slab with an intercalated A$_2$Z$_2$ layer, yielding high thermodynamic stability and intrinsic 
two-dimensionality. This architecture supports extensive chemical flexibility, with over 70 predicted 
stable monolayers across multiple structural symmetries ($\alpha$, $\beta$, $\gamma$, $\delta$, 
see few examples in Fig.~\ref{fig:struc}) and electronic 
ground states. The MA$_2$Z$_4$ family hosts a wide range of electronic phases including semiconductors, 
metals, cold metals, spin-gapped metals, spin-gapless semiconductors, ferromagnets, and topological 
insulators, and exhibits properties such as ultrahigh carrier mobilities, large exciton binding energies,
strain-tunable magnetism, and valley-selective optical responses \cite{Wang,latychevskaia2025mosi2n4,tho2023ma2z4,Guo2024,Otrokov2019,Gao2024_AFMdiode}. 
Given this chemical and electronic diversity, a quantitative study of Coulomb interaction screening, correlation 
strength, and magnetic tendencies across the MA$_2$Z$_4$ family is both timely and essential for future 
theoretical and applied advances, particularly in the design of correlated 2D materials for spintronic, 
optoelectronic, and quantum information technologies.

In this work, we present a comprehensive first-principles investigation of screening behavior, 
Coulomb interaction parameters, correlation strength, and collective excitations across more 
than 60 MA$_2$Z$_4$ monolayers using the constrained random phase approximation (cRPA). We compute 
effective (partially screened) on-site interactions ($U$, $U'$, and $J$) and their spatial 
dependence across different correlated subspaces defined by orbital character and local symmetry. 
In semiconducting compounds, we find that long-range nonlocal interactions persist, suggesting 
unconventional screening behavior and the relevance of strong excitonic effects beyond simple 
dielectric screening. In cold-metallic systems, sizable long-range interactions are found despite 
the presence of free carriers, highlighting their atypical metallic screening. Among 33-valence-electron 
systems, our analysis of on-site and off-site interactions in the $\beta_2$ phase reveals candidate 
materials for charge-density-wave or Mott instabilities, with $U_{\mathrm{eff}} > W$ in multiple 
cases. Several V- and Nb-based compounds exhibit intermediate-to-strong correlation strength ($U/W > 1$), 
and spin-polarized density-functional theory (DFT)  calculations show finite magnetic moments. 
Applying the Stoner criterion with cRPA-derived Stoner parameters, we identify magnetic instabilities 
in various V-, Cr-, Nb-, and Mn-based compounds. Finally, for selected cold-metallic materials, 
we compute plasmon dispersions and observe deviations from the conventional $\sqrt{q}$ behavior, 
indicating the emergence of slow, nearly non-dispersive plasmon modes. Although explicit many-body 
corrections beyond DFT are not included, our results offer a rigorous, material-specific foundation 
for future DFT+DMFT or model Hamiltonian studies. The comprehensive dataset presented here serves
as a reference for understanding correlation-driven phenomena and guiding the design of MA$_2$Z$_4$-based 
materials for applications in spintronics, excitonics, and plasmonics.

\begin{figure*}[tp]
\centering
\includegraphics[width=0.96\linewidth]{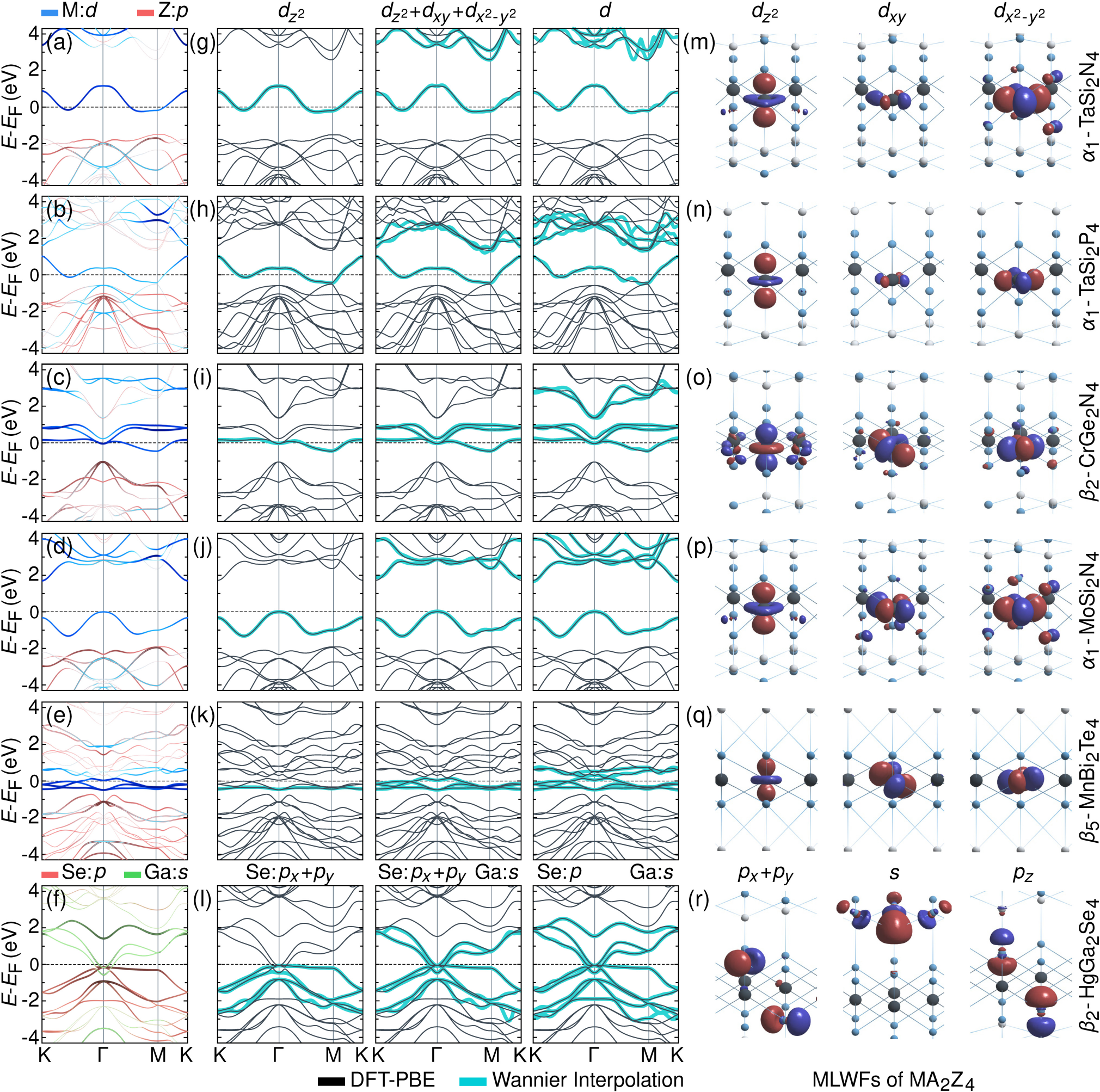}
\vspace{-0.1 cm}
\caption{Electronic structure and maximally localized Wannier functions (MLWFs) for representative 
MA$_2$Z$_4$ compounds. Panels (a)--(f) show the projected DFT-PBE band structures of $\alpha_1$-TaSi$_2$N$_4$, 
$\alpha_1$-TaSi$_2$P$_4$, $\beta_2$-CrGe$_2$N$_4$, $\alpha_1$-MoSi$_2$N$_4$, $\beta_5$-MnBi$_2$Te$_4$, and 
$\beta_2$-HgGa$_2$Se$_4$. Panels (g)--(l) compare the original DFT-PBE bands (gray) with the Wannier-interpolated
bands (cyan) for three different correlated subspaces. Panels (m)--(r) display the spatial distributions of selected 
MLWFs used in the construction of these subspaces. Left column: $d_{z^2}$-like MLWF (one-orbital model). Middle column: 
$d_{xy}$-like MLWF from a three-orbital subspace ($d_{z^2}$ + $d_{xy}$ + $d_{x^2-y^2}$). Right column: $d_{x^2-y^2}$-like 
MLWF from the full $d$-manifold. For the topological compound $\beta_2$-HgGa$_2$Te$_4$, the relevant subspaces consist
of $p_x$+$p_y$, $s$ + $p_x$+$p_y$, and $s$ + $p$ orbitals.}
\label{fig:interpol}
\end{figure*}

\section*{Results and Discussion}

The structural and electronic diversity of MA$_2$Z$_4$ monolayers, widely reported in the literature, 
motivates our systematic investigation of Coulomb interaction parameters and magnetic tendencies across this family.
While all compounds share a common septuple-layer architecture, differences in atomic arrangement and symmetry 
give rise to multiple stable polytypes. Figure~\ref{fig:struc} shows representative low-energy configurations 
across the $\alpha_1$, $\alpha_2$, $\beta_1$, $\beta_2$, and $\beta_5$ structural classes, as identified in 
high-throughput studies and confirmed by phonon calculations~\cite{Wang}. Each phase exhibits 
a distinct coordination environment of the transition-metal atom, ranging from trigonal prismatic (1H-like) 
to ideal octahedral (1T-like), which modulates the crystal field splitting and influences the definition 
of the correlated subspace. We also include a representative charge-density-wave (CDW) reconstructed $\beta_2$ 
structure to highlight possible symmetry-lowering effects in strongly correlated regimes. These structural 
motifs form the basis for selecting correlated orbitals in the subsequent cRPA analysis.

\subsection*{Correlated subspaces in 2D MA$_2$Z$_4$ materials}

To compute screened Coulomb interactions within the cRPA framework, it is essential to define an 
appropriate correlated subspace for each material. We identify this subspace based on the dominant
orbital character near the Fermi level, as obtained from DFT calculations 
and maximally localized Wannier function (MLWF) analysis. Across the majority of MA$_2$Z$_4$ compounds, 
the low-energy electronic structure is primarily derived from the transition metal (M) $d$ orbitals, 
with negligible contribution from A or Z atoms. This trend enables a consistent treatment using 
$d$-orbital-based correlated subspaces. Depending on the degree of band separation and orbital mixing, 
we construct MLWFs for one-, three-, and five-orbital models. Figure~\ref{fig:interpol} illustrates the 
comparison between DFT and Wannier-interpolated bands, along with representative MLWF spatial profiles. 
Additional examples of Wannier interpolations and orbital-resolved character for different subspaces 
are provided in Supplementary Figures S1–S3.

For semiconductors and cold metallic systems with isolated $d_{z^2}$ bands crossing the Fermi level, such as
$\alpha_1$-TaSi$_2$N$_4$ and $\alpha_1$-NbSi$2$P$4$, a one-orbital model is sufficient. In materials where 
additional $d_{xy}$ and $d_{x^2-y^2}$ states contribute within a narrow energy window (e.g., $\alpha_1$-MoSi$_2$N$_4$), 
a minimal three-orbital model offers a more accurate description. In more complex cases like $\beta_2$-CrGe$_2$N$4$, 
where strong hybridization between $d_{z^2}$ and other $e_g$ orbitals occurs, the correlated subspace must be extended 
to include all five $d$ orbitals. These subspace definitions enable a consistent treatment of electron correlation 
effects across the $d$-dominated MA$_2$Z$_4$ materials.

A notable exception to this $d$-orbital-based description arises in the 32-valence-electron MA$_2$Z$_4$ compounds. 
These systems, typically involving group-12 transition metals (e.g., Zn, Cd, Hg) combined with lighter A 
and Z atoms, feature low-energy states near the Fermi level that originate primarily from $s$ and $p$ orbitals 
of the A and Z atoms, rather than from M-$d$ states. For these $s/p$-dominated semiconductors, we construct
two-, three-, and four-orbital correlated subspaces based on appropriate combinations of $s$ and $p$ orbitals. 
These models—illustrated in Fig.~\ref{fig:interpol}(f), (l), and (r)—are essential for accurately capturing 
correlation effects in non-$d$-like semiconducting phases.

The choice of correlated subspace influences the calculated interaction parameters and should reflect the physical 
properties of interest. Minimal models are sufficient for evaluating low-energy transport properties, 
while more complete subspaces are required for describing optical excitations or many-body instabilities. In the 
following sections, we report Coulomb interaction parameters for each material using one-orbital ($d_{z^2}$), 
three-orbital ($d_{z^2} + d_{xy} + d_{x^2 - y^2}$ or $d_{z^2} + e_g$), and five-orbital ($d$-manifold) models, 
as well as $s$/$p$-based subspaces for the 32-valence-electron materials.

\subsection*{Coulomb interactions in 2D MA$_2$Z$_4$ with 32 and 34 valence electrons}

We begin our quantitative analysis of Coulomb interactions with the 34-valence-electron MA$_2$Z$_4$ monolayers. This
subset comprises a large number of semiconducting compounds based on Mo and W, along with a smaller set of Cr-based 
materials that display metallic or magnetic ground states. These systems are particularly relevant for understanding 
electronic correlation effects, as their low-energy states are dominated by transition-metal $d$ orbitals. The 
prototypical MoSi$_2$N$_4$, the first experimentally synthesized member of the MA$_2$Z$_4$ family, also belongs to this 
class, making it a natural reference point for our analysis~\cite{hong2020chemically}. While on-site Coulomb interactions 
play a central role in local correlation physics, the long-range behavior of the interaction is equally crucial in 2D materials,
where screening is reduced and non-local effects can become significant. To ensure consistency in the definition of 
interaction parameters and to simplify the construction of localized Wannier functions, all cRPA and RPA calculations 
are performed using a non-spin-polarized state, even for compounds that exhibit magnetic ordering in spin-polarized 
calculations. This choice preserves spin symmetry, avoids complications related to spin-dependent screening channels, 
and prevents spurious screening effects introduced by broken-symmetry DFT states. It also aligns with the formal 
structure of many-body techniques such as DFT+DMFT or model Hamiltonian approaches, in which magnetic fluctuations 
are treated dynamically and self-consistently within the correlated subspace.

\begin{figure*}[tp]
\centering
\includegraphics[width=1.0\linewidth]{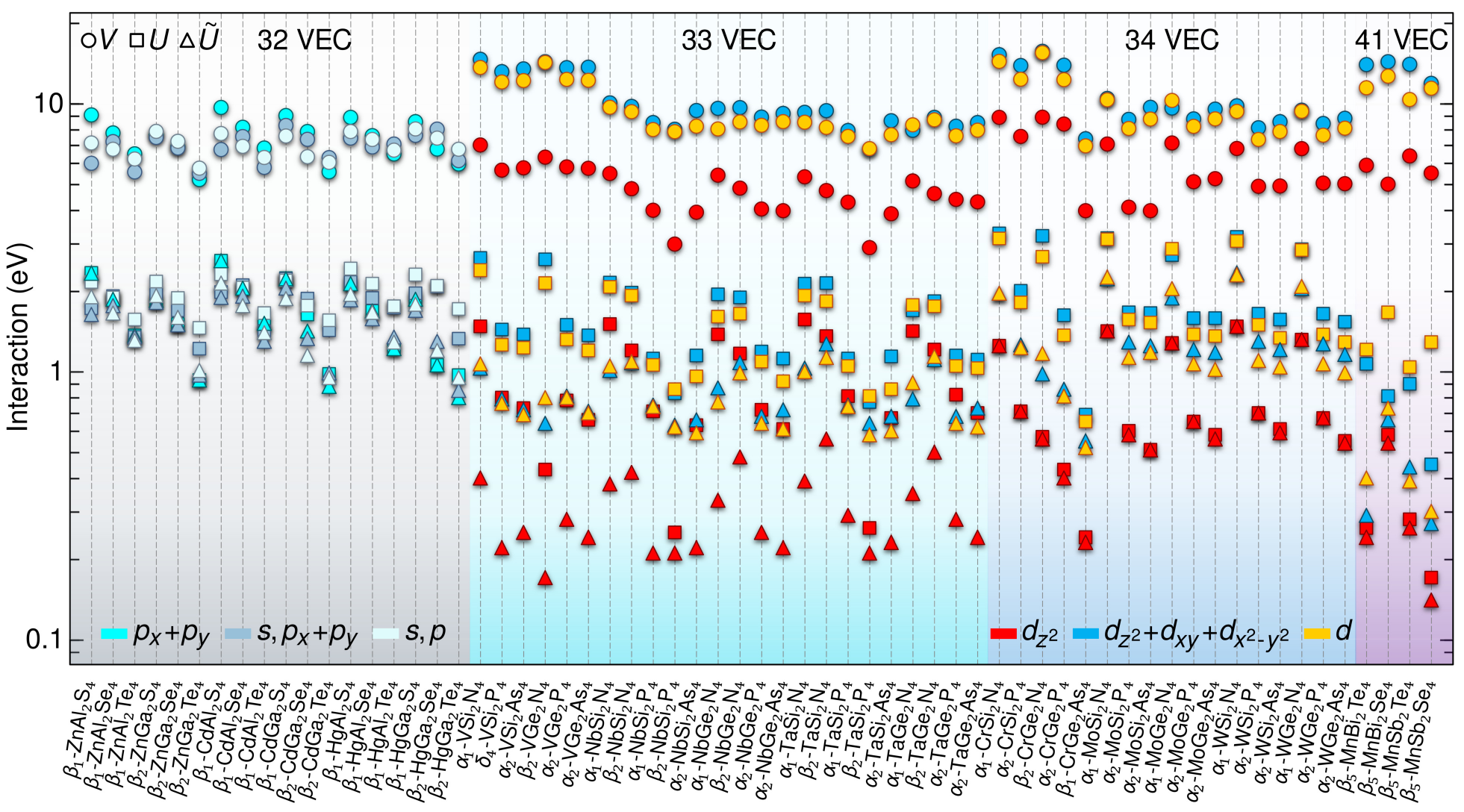}
\caption{On-site intra-orbital Coulomb interaction parameters — bare interaction $V$, partially screened interaction 
$U$, and fully screened interaction $\tilde{U}$ — for 2D MA$_2$Z$_4$ compounds with 32, 33, 34, and 41 
valence electrons. Results are shown for different correlated subspaces, enabling comparison of the strength and 
screening of electronic interactions across a wide range of chemical compositions and electronic configurations. 
The abbreviation VEC refers to valence electron count.}
\label{fig:U}
\end{figure*}

As a first step, we examine the bare intra-orbital Coulomb interaction \( V \), which provides a measure of the 
spatial localization of the underlying Wannier orbitals. Table~\ref{table1} and Fig.~\ref{fig:U} show that \( V \) 
varies substantially across the MA$_2$Z$_4$ family, spanning a range from approximately 4 to 16~eV depending on 
chemical composition, lattice structure, and the choice of correlated subspace. Across all compounds, we find that 
Cr-based systems exhibit the largest \( V \) values, consistent with the stronger localization of 3$d$ orbitals 
relative to their 4$d$ (Mo) and 5$d$ (W) counterparts. In parallel, a clear anion trend is observed: for a fixed 
transition metal and cation, the bare interaction decreases as the anion changes from N to P to As. This trend 
reflects the increased lattice spacing and weaker hybridization associated with heavier anions, both of which 
contribute to more delocalized Wannier functions. The influence of cation substitution (Si vs.\ Ge) is more subtle
but generally results in a modest reduction of \( V \), likely due to lattice expansion and altered bonding geometry.
Overall, the observed variations in \( V \) highlight the strong sensitivity of orbital localization to both electronic 
and structural degrees of freedom.

\begin{table*}[tp]
\caption{Crystal symmetry, lattice parameter, electronic ground state, correlated subspace, and Coulomb interaction 
parameters for $d$ orbitals of transition-metal atoms in 2D MA$_2$Z$_4$ compounds with 34 valence electrons
(M = Cr, Mo, W; A = Si, Ge; Z = N, P, As). Listed are the bare intra-orbital interaction $V$, the partially screened 
Hubbard–Kanamori parameters [$U$, $U^{\prime}$, $J$], and the fully screened counterparts [$\tilde{U}$, $\tilde{U}^{\prime}$,  
$\tilde{J}$]. For each compound, results are provided for three different 
correlated subspaces: a single-orbital model ($d_{z^2}$), a three-orbital model ($d_{z^2}$ + $d_{xy}$ + $d_{x^2 - y^2}$, 
or $d_{z^2}$ + $e_{g}$), and the full $d$-orbital manifold. Lattice parameters are taken from Ref.~\onlinecite{Wang}. 
Abbreviations: SC — semiconductor; SGM — spin-gapped metal; FM — ferromagnet.}
\begin{ruledtabular}
 \begin{tabular}{lccccccccccc}
\multirow{2}{*}{MA$_2$Z$_4$} & \multirow{2}{*}{Phase} & a & Ground & Correlated & $V$ & $U$ & $U^{\prime}$ & $J$ & $\tilde{U}$ & $\tilde{U}^{\prime}$ & $\tilde{J}$ \\
	&  & (\AA) & State & Subspace & (eV) & (eV) & (eV) & (eV) & (eV) & (eV) & (eV) \\ \hline
			\multirow{3}{*}{CrSi$_2$N$_4$} & \multirow{3}{*}{$\alpha_{1}$} & \multirow{3}{*}{2.84} & \multirow{3}{*}{SC} & $ d_{z^2}$ & 9.00 & 1.25 & - & - & 1.25 & - & -  \\
			&  &  &  & $ d_{z^2}$+$ d_{xy}$+$ d_{x^2-y^2}$ & 15.37 & 3.28 & 2.40 & 0.48 & 1.97 & 1.31 & 0.35 \\
			&  &  &  & $d$ & 14.52 & 3.15 & 2.54 & 0.39 & 2.00 & 1.42 & 0.30 \\
			\multirow{3}{*}{CrSi$_2$P$_4$} & \multirow{3}{*}{$\alpha_{2}$} & \multirow{3}{*}{3.41} & \multirow{3}{*}{SC} & $ d_{z^2}$ & 7.60 & 0.71 & - & - & 0.71 & - & - \\
			&  &  &  & $ d_{z^2}$+ $ d_{xy}$+$ d_{x^2-y^2}$ & 14.00 & 2.01 & 1.27 & 0.38 & 1.26& 0.72 & 0.28 \\
			&  &  &  & $d$ & 12.44 & 1.82 & 1.33 & 0.30 & 1.23 & 0.78 & 0.24 \\
               \multirow{3}{*}{CrSi$_2$As$_4$} & \multirow{3}{*}{$\beta_{1}$} & \multirow{3}{*}{3.68} & \multirow{3}{*}{FM} & $ d_{z^2}$ & 5.81 & 0.26 & - & - & 0.26 & - & - \\
               &  &  &  & $ d_{z^2}$+$ e_{g}$ & 13.09 & 1.10 & 0.49 & 0.30 & 0.61 & 0.23 & 0.18 \\
               &  &  &  & $d$ & 11.96 & 1.07 & 0.65 & 0.25 & 0.63 & 0.29 & 0.17 \\
			\multirow{3}{*}{CrGe$_2$N$_4$} & \multirow{3}{*}{$\beta_{2}$} & \multirow{3}{*}{3.06} & \multirow{3}{*}{SGM} & $ d_{z^2}$ & 8.96 & 0.57 & - & - & 0.57 & - & - \\
			&  &  &  & $ d_{z^2}$+$ e_{g}$ & 15.83  & 3.22 & 2.34 & 0.44 & 0.98 & 0.49 & 0.20  \\
			&  &  &  & $d$ & 15.64 & 2.69 & 1.96 & 0.42 & 1.17 & 0.54 & 0.24 \\
			\multirow{3}{*}{CrGe$_2$P$_4$} & \multirow{3}{*}{$\alpha_{2}$} & \multirow{3}{*}{3.49} & \multirow{3}{*}{SC} & $ d_{z^2} $ & 7.49 & 0.65 & - & - & 0.63 & - & -  \\
			&  &  &  & $ d_{z^2}$+$ d_{xy}$+$ d_{x^2-y^2}$ & 14.11 & 1.91 & 1.20 & 0.36 & 1.14 & 0.64 & 0.26 \\
			&  &  &  & $d$ & 12.47 & 1.73 & 1.27 & 0.27 & 1.10 & 0.69 & 0.22 \\
			\multirow{3}{*}{CrGe$_2$As$_4$} & \multirow{3}{*}{$\beta_{1}$} & \multirow{3}{*}{3.78} & \multirow{3}{*}{FM} & $ d_{z^2}$ & 4.00 & 0.24 & - & - & 0.23 & - & - \\
			&  &  &  & $d_{z^2}$+$e_{g}$ & 7.45 & 0.70 & 0.33 & 0.10 & 0.55 & 0.25 & 0.09 \\
			&  &  &  & $d$ & 7.00 & 0.65 & 0.35 & 0.08 & 0.51 & 0.26 & 0.07 \\
			\multirow{3}{*}{MoSi$_2$N$_4$} & \multirow{3}{*}{$\alpha_{1}$} & \multirow{3}{*}{2.91} & \multirow{3}{*}{SC} & $ d_{z^2}$ &7.11 & 1.42 & - & - & 1.42 & - & - \\
			&  &  &  & $ d_{z^2}$+$ d_{xy}$ + $ d_{x^2-y^2}$ & 10.54 & 3.16 & 2.58 & 0.33 & 2.23 & 1.71 & 0.29 \\
			&  &  &  & $d$ & 10.41 & 3.12 & 2.69 & 0.27 & 2.26 & 1.80 & 0.25 \\
			\multirow{3}{*}{MoSi$_2$P$_4$} & \multirow{3}{*}{$\alpha_{2}$} & \multirow{3}{*}{3.46} & \multirow{3}{*}{SC} & $d_{z^2}$ & 4.13 & 0.60 & - & - & 0.58 & - & - \\
			&  &  &  & $ d_{z^2}$+$ d_{xy}$+$ d_{x^2-y^2}$ & 8.80 & 1.67 & 1.23 & 0.24 & 1.29 & 0.88 & 0.21 \\
			&  &  &  & d & 8.13 & 1.57 & 1.26 & 0.19 & 1.13 & 0.94 & 0.12 \\
			\multirow{3}{*}{MoSi$_2$As$_4$} & \multirow{3}{*}{$\alpha_{2}$} & \multirow{3}{*}{3.61} & \multirow{3}{*}{SC} & $ d_{z^2}$ & 4.00 & 0.51 & - & - & 0.51 & - & - \\
			&  &  &  & $ d_{z^2}$+$ d_{xy}$+$ d_{x^2-y^2}$ & 9.77 & 1.66 & 1.14 & 0.28 & 1.25 & 0.78 & 0.24 \\
			&  &  &  & $d$ & 8.82 & 1.53 & 1.17 & 0.22 & 1.18 & 0.83 & 0.20 \\	
			\multirow{3}{*}{MoGe$_2$N$_4$} & \multirow{3}{*}{$\alpha_{1}$} & \multirow{3}{*}{3.02} & \multirow{3}{*}{SC} & $ d_{z^2}$ & 7.18 & 1.28 & - & - & 1.28 & - & - \\
			&  &  &  & $ d_{z^2}$+$ d_{xy}$+$ d_{x^2-y^2}$ & 9.70 & 2.73 & 2.21 & 0.29 & 1.89 & 1.43 & 0.25 \\
			&  &  &  & $d$ & 10.36 & 2.88 & 2.45 & 0.27 & 2.05 & 1.59 & 0.24 \\
			\multirow{3}{*}{MoGe$_2$P$_4$} & \multirow{3}{*}{$\alpha_{2}$} & \multirow{3}{*}{3.53} & \multirow{3}{*}{SC} & $ d_{z^2} $ & 5.14 & 0.65 & - & - & 0.65 & - & -  \\
			&  &  &  & $ d_{z^2}$+$ d_{xy}$+$ d_{x^2-y^2}$ & 8.84 & 1.59 & 1.14 & 0.23 & 1.21 & 0.81 & 0.20 \\
			&  &  &  & $d$ & 8.28 & 1.38 & 1.18 & 0.12 & 1.07 & 0.88 & 0.11 \\
			\multirow{3}{*}{MoGe$_2$As$_4$} & \multirow{3}{*}{$\alpha_{2}$} & \multirow{3}{*}{3.69} & \multirow{3}{*}{SC} & $ d_{z^2}$ & 5.28 & 0.58 & - & - & 0.56 & - & - \\
			&  &  &  & $d_{z^2}$+$d_{xy}$+$d_{x^2-y^2}$ & 9.61 & 1.59 & 1.08 & 0.27 & 1.18 & 0.72 & 0.23 \\
			&  &  &  & $d$ & 8.84 & 1.36 & 1.13 & 0.14 & 1.02 & 0.79 & 0.13 \\
			\multirow{3}{*}{WSi$_2$N$_4$} & \multirow{3}{*}{$\alpha_{1}$} & \multirow{3}{*}{2.91} & \multirow{3}{*}{SC} & $d_{z^2}$ & 6.85 & 1.48 & - & - & 1.48 & - & - \\
			&  &  &  & $d_{z^2}$+$d_{xy}$+$d_{x^2-y^2}$ & 9.90 & 3.19 & 2.65 & 0.33 & 2.34 & 1.83 & 0.29 \\
			&  &  &  & $d$ & 9.43 & 3.08 & 2.69 & 0.27 & 2.31 & 1.89 & 0.25 \\
			\multirow{3}{*}{WSi$_2$P$_4$} & \multirow{3}{*}{$\alpha_{2}$} & \multirow{3}{*}{3.46} & \multirow{3}{*}{SC} & $ d_{z^2}$ & 4.94 & 0.70 & - & - & 0.70 & - & -  \\
			&  &  &  & $ d_{z^2}$+$ d_{xy}$+$ d_{x^2-y^2}$ & 8.19 & 1.65 & 1.24 & 0.23 & 1.30 & 0.91 & 0.21 \\
			&  &  &  & $d$ & 7.40 & 1.50 & 1.21 & 0.18 & 1.10 & 0.94 & 0.10 \\
			\multirow{3}{*}{WSi$_2$As$_4$} & \multirow{3}{*}{$\alpha_{2}$} & \multirow{3}{*}{3.61} & \multirow{3}{*}{SC} & $ d_{z^2}$ & 4.95 & 0.61 & - & - & 0.59 & - & - \\
			&  &  &  & $ d_{z^2}$+$ d_{xy}$+$d_{x^2-y^2}$ & 8.63 & 1.57 & 1.12 & 0.25 & 1.21 & 0.80 & 0.22 \\
			&  &  &  & $d$ & 7.92 & 1.34 & 1.16 & 0.12 & 1.04 & 0.86 & 0.11 \\
			\multirow{3}{*}{WGe$_2$N$_4$} & \multirow{3}{*}{$\alpha_{1}$} & \multirow{3}{*}{3.02} & \multirow{3}{*}{SC} & $ d_{z^2}$ & 6.84 & 1.32 & - & - & 1.32 & - & - \\
			&  &  &  & $ d_{z^2}$+$ d_{xy}$+$ d_{x^2-y^2}$ & 9.53 & 2.87 & 2.33 & 0.30 & 2.05 & 1.55 & 0.26 \\
			&  &  &  & $d$ & 9.44 & 2.85 & 2.44 & 0.26 & 2.09 & 1.65 & 0.24 \\
			\multirow{3}{*}{WGe$_2$P$_4$} & \multirow{3}{*}{$\alpha_{2}$} & \multirow{3}{*}{3.54} & \multirow{3}{*}{SC} & $ d_{z^2}$ & 5.08 & 0.67 & - & - & 0.67 & - & - \\
			&  &  &  & $ d_{z^2}$+$ d_{xy}$+$ d_{x^2-y^2}$ & 8.50 & 1.65 & 1.19 & 0.23 & 1.27 & 0.85 & 0.21 \\
			&  &  &  & $d$ & 7.68 & 1.38 & 1.21 & 0.11 & 1.07 & 0.90 & 0.10 \\
			\multirow{3}{*}{WGe$_2$As$_4$} & \multirow{3}{*}{$\alpha_{2}$} & \multirow{3}{*}{3.69} & \multirow{3}{*}{SC} & $ d_{z^2}$ & 5.05 & 0.55 & - & - & 0.54 & - & - \\
			&  &  &  & $ d_{z^2}$+$ d_{xy}$+$ d_{x^2-y^2}$ & 8.88 & 1.54 & 1.05 & 0.25 & 1.16 & 0.72 & 0.22 \\
			&  &  &  & $d$ & 8.16 & 1.30 & 1.10 & 0.12 & 0.99 & 0.79 & 0.11 \\
		\end{tabular} 
		\label{table1}
	\end{ruledtabular}
\end{table*}

In addition to these chemical trends, \( V \) exhibits a systematic dependence on the size of the 
correlated subspace. In particular, the one-orbital (\( d_{z^2} \)) model yields the lowest \( V \), 
the three-orbital (\( d_{z^2} \), \( d_{xy} \), and \( d_{x^2 - y^2} \)) model gives a significantly 
higher value, and the five-orbital ($d$) model produces a \( V \) that is slightly smaller than in 
the three-orbital case. The reduced \( V \) in the 1-orbital subspace reflects the strong delocalization 
of the isolated Wannier orbital, which is more extended due to the lack of nearby states within the 
same manifold. Conversely, the slight drop in \( V \) for the five-orbital case relative to the 
three-orbital model can be attributed to enhanced delocalization arising from metal--ligand ($p$--$d$) 
hybridization. This consistent trend across the 34-electron MA$_2$Z$_4$ series reveals how the 
choice of correlated subspace influences Wannier function localization and, consequently, the strength 
of the bare Coulomb 
interaction \( V \).

To quantify the effects of screening, we compute the effective (partially screened) Coulomb interaction parameters, namely 
the intra-orbital Hubbard interaction \( U \), the inter-orbital interaction \( U' \), and the exchange interaction \( J \), 
within the constrained random phase approximation (cRPA), which excludes screening contributions from the low-energy correlated 
subspace. Additionally, we evaluate the fully screened interactions \( \tilde{U} \), \( \tilde{U}' \), and \( \tilde{J} \) 
using the full random phase approximation (RPA), which includes all electronic transitions. Table~\ref{table1} provides a 
comprehensive dataset of Coulomb parameters for all 34-valence-electron MA$_2$Z$_4$ compounds, evaluated for three different 
correlated subspaces. These results are further contextualized in Fig.~\ref{fig:U}, which displays the intra-orbital 
\( V \), \( U \), and \( \tilde{U} \) values not only for the 34-electron subset but also across the broader set of 
MA$_2$Z$_4$ compounds with 32, 33, and 41 valence electrons. The full \( 3 \times 3 \) and \( 5 \times 5 \) cRPA Coulomb 
matrices for all 34-electron compounds are provided in the Supplementary Material, enabling more detailed many-body 
modeling beyond the averaged values discussed here.

In the semiconducting Mo- and W-based compounds, the partially screened Hubbard-Kanamori interaction \( U \) ranges from 0.5 to 3.2~eV, 
depending on the correlated subspace and chemical environment. These values are comparable to or larger than those reported 
for other 2D semiconductors such as MoS$_2$~\cite{Ramezani,yekta_2021}, consistent with reduced dielectric screening 
in low-dimensional systems. Notably, \( U \) increases along the transition-metal series from Cr to Mo to W, reflecting the 
interplay between orbital character and screening efficiency. This trend is indicative of weaker $p$-to-$d$ screening in heavier 
transition metals, where larger band gaps reduce the effectiveness of polarization screening from ligand states. For example, 
in the full \( d \)-orbital subspace, the interaction \( U \) increases from 1.38~eV in MoGe$_2$P$_4$ to 2.85~eV in WGe$_2$N$_4$. 
Anion substitution also plays a significant role: within the MoSi$_2$Z$_4$ series, the interaction \( U \) decreases from 
3.12~eV in MoSi$_2$N$_4$ to 1.57~eV in MoSi$_2$P$_4$, and further to 1.53~eV in MoSi$_2$As$_4$, reflecting the enhanced screening 
in compounds with heavier anions and wider lattice constants. A similar trend is observed in the W-based series, with \( U \) in 
the full-\( d \) subspace decreasing from 3.08~eV in WSi$_2$N$_4$ to 1.50~eV in WSi$_2$P$_4$, and 1.34~eV in WSi$_2$As$_4$. 
Comparing Si- and Ge-based compounds reveals a more nuanced effect: while lattice expansion associated with Ge substitution 
tends to slightly reduce \( U \), the trend is not always monotonic. For instance, WSi$_2$N$_4$ and WGe$_2$N$_4$ exhibit similar 
values of \( U \) (3.08~eV vs. 2.85~eV), suggesting that structural and hybridization effects can partially compensate for the 
expected increase in screening. Overall, these trends reveal the subtle interplay between chemical composition, orbital 
character, and lattice geometry in determining the effective strength of electron-electron interactions in the MA$_2$Z$_4$ family.

\begin{figure*}[tp] 	
\centering
\includegraphics[width=1.0\linewidth]{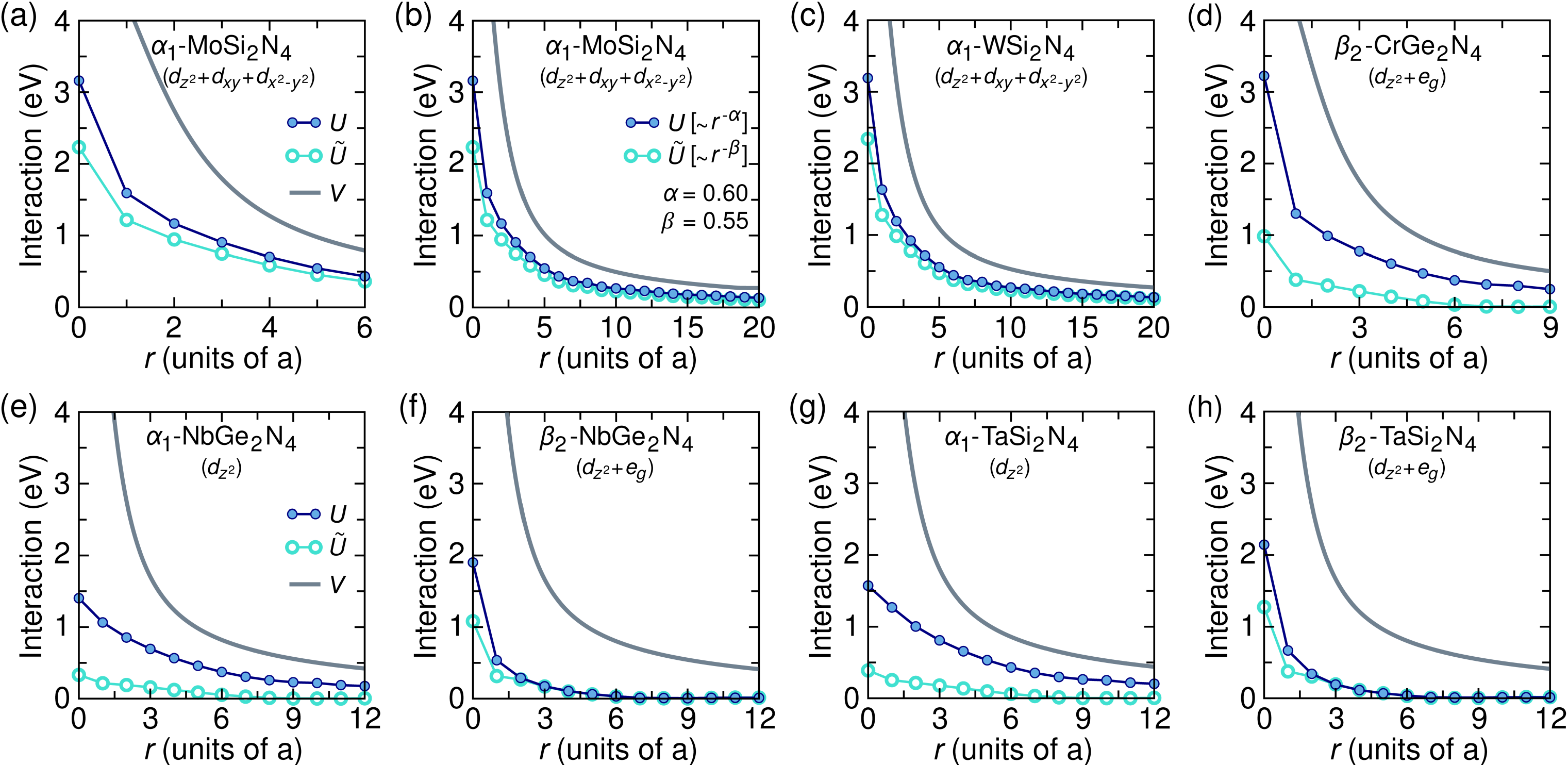}
\centering
\vspace{-0.5 cm}
\caption{Real-space decay of the intra-orbital Coulomb interaction for selected MA$_2$Z$_4$ compounds with 34 and 33 valence electrons. 
Shown are the bare interaction $V(r)$ (solid lines), the partially screened interaction $U(r)$, and the fully screened interaction $\tilde{U}(r)$ 
as functions of distance $r$, expressed in units of the in-plane lattice constant $a$. All results correspond to the $d_{z^2}$+$d_{xy}$+$d_{x^2-y^2}$ 
(or $d_{z^2}$+$e_g$) correlated subspace. Panels (a)--(d) present 34-valence-electron compounds: (a) displays $V(r)$, $U(r)$, and $\tilde{U}(r)$ for 
$\alpha_1$-MoSi$_2$N$_4$ up to $r = 6a$;  (b) extends this analysis to $r = 20a$ and includes power-law fits, where $U(r) \sim r^{-\alpha}$ and $\tilde{U}(r) 
\sim r^{-\beta}$ with $\alpha = 0.6$ and $\beta = 0.55$, respectively;  (c) shows the analogous decay for $\alpha_1$-WSi$_2$N$_4$ up to $r = 20a$, 
exhibiting a similar slow screening behavior without fitted exponents;  (d) presents $\beta_2$-CrGe$_2$N$_4$ up to $r = 9a$, where the faster decay 
of $\tilde{U}(r)$ reflects stronger metallic screening. Panels (e)--(h) extend the analysis to 33-valence-electron compounds: 
(e) and (f) show the decay of $V(r)$, $U(r)$, and $\tilde{U}(r)$ up to $r=12a$ for $\alpha_1$- and $\beta_2$-NbGe$_2$N$_4$, respectively; 
(g) and (h) present the corresponding results for $\alpha_1$- and $\beta_2$-TaSi$_2$N$_4$ up to $r=12a$.}
\label{fig:long-MoW}
\end{figure*}

To gain deeper insight into the nature of screening in MA$_2$Z$_4$ materials, we extend our analysis beyond the 
on-site limit and examine the real-space decay of the intra-orbital Coulomb interaction \( U(r) \). Understanding 
the spatial profile of \( U(r) \) is essential for assessing the range and strength of electron-electron 
interactions, particularly in 2D systems where non-local screening effects are pronounced. 
In the following, we focus on representative compounds from the 34-valence-electron family to explore how 
screening behavior evolves with electronic structure, dimensionality, and material composition. These results 
reveal substantial differences between semiconducting and metallic systems. Complete long-range interaction data, 
including \( U(r) \) values up to the sixth nearest neighbor for all 34-electron compounds and all correlated subspaces, 
are provided in Table S1 in the Supplementary Material.

The long-range behavior of the Coulomb interaction is particularly revealing in the case of semiconductors, where reduced 
dimensionality suppresses conventional dielectric screening. As shown in Fig.~\ref{fig:long-MoW}, we plot the real-space 
decay of the bare, partially screened, and fully screened intra-orbital Coulomb interactions—$V(r)$, $U(r)$, and 
$\tilde{U}(r)$—for the representative compounds $\alpha_1$-MoSi$_2$N$_4$, $\alpha_1$-WSi$_2$N$_4$, and spin-gapped metal
$\beta_2$-CrGe$_2$N$_4$. In all cases, $r$ is expressed in units of the in-plane lattice constant $a$. For the 
semiconducting systems, we observe a pronounced deviation from conventional metallic or bulk-like screening behavior. 
In $\alpha_1$-MoSi$_2$N$_4$, shown in panel (a), the effective interaction $U(r)$ remains substantial up to at least $6a$, 
and its long-range tail persists even at $20a$ [panel (b)]. A power-law fit reveals that $U(r) \sim r^{-0.6}$ and 
$\tilde{U}(r) \sim r^{-0.55}$, in stark contrast to the bare $V(r) \sim 1/r$ decay. This unusually slow screening reveals 
the nonlocal character of Coulomb interactions in 2D semiconductors.

A similar trend is found for $\alpha_1$-WSi$_2$N$_4$ [panel (c)], where $U(r)$ and $\tilde{U}(r)$ exhibit nearly identical
decay profiles to those in MoSi$_2$N$_4$, albeit without fitted exponents. The persistence of long-range interactions in 
both cases reveals a key feature of semiconducting MA$_2$Z$_4$ systems: the dielectric screening is strongly distance-dependent
and cannot be captured by a simple static dielectric constant $\epsilon$. For example, in $\alpha_1$-MoSi$_2$N$_4$, the 
effective dielectric ratio $\epsilon(r) = V(r)/U(r)$ decreases with increasing distance, taking values of $\epsilon(a) 
\approx 2.8$, $\epsilon(2a) \approx 2.1$, and $\epsilon(3a) \approx 1.8$. This non-monotonic screening response indicates 
that long-range interactions are weakly screened and that $U(r)$ and $\tilde{U}(r)$ do not follow a conventional $1/\epsilon r$ form.

In contrast, the spin-gapped metal $\beta_2$-CrGe$_2$N$_4$ [panel (d)] shows a much more localized interaction profile. Here, 
the screened Coulomb interaction decays rapidly and becomes negligible beyond approximately $6a$, reflecting the stronger 
screening expected in metallic systems. This behavior is consistent with the emergence of delocalized carriers that effectively
suppress long-range components of the Coulomb potential. Extended data for other metallic compounds, including
off-site $U(r)$ values up to $6a$, are provided in Table I of the Supplementary Material.

Overall, these results highlight the unconventional nature of screening in semiconducting MA$_2$Z$_4$ monolayers. The persistence 
of nonlocal Coulomb interactions over several lattice spacings, as seen in both the partially screened \( U(r) \) and fully screened 
\( \tilde{U}(r) \), has important implications for electronic and optical properties. In particular, the long-range behavior of 
\( \tilde{U}(r) \) provides a microscopic basis for the large exciton binding energies and non-Rydberg excitonic series 
reported in MoSi$_2$N$_4$ and related materials~\cite{hong2020chemically,Wozniak,Sun}. While excitonic properties are not computed in this work, 
our analysis of distance-dependent screening establishes key input parameters for many-body calculations and supports the predictive 
design of MA$_2$Z$_4$-based optoelectronic and excitonic devices.

In light of these findings, the calculated off-site Coulomb interaction values, presented in Table I of the Supplementary 
Material, offer a valuable foundation for first-principles methods that go beyond purely local correlation models. 
In particular, the DFT+$U$+$V$ framework, which supplements on-site Hubbard interactions with non-local intersite terms $V$, 
has proven effective in describing semiconductors with strong covalent bonding~\cite{Campo-2010,Timrov-2021}. Given the 
covalently bonded in-plane structure of MA$_2$Z$_4$ monolayers and the persistence of long-range interactions observed
in our calculations, especially for semiconducting compounds, DFT+$U$+$V$ emerges as a natural candidate for improved electronic 
structure modeling in this class of materials. The distance-resolved $U(r)$ values provided here can serve as physically 
motivated input parameters for such extended Hubbard models, enabling future studies to better capture the interplay between 
short- and long-range correlation effects in 2D semiconductors.

While the primary focus of this work has been on 34-valence-electron MA$_2$Z$_4$ monolayers with $d$-orbital-dominated 
correlated subspaces, it is also important to consider the complementary class of materials with 32 valence electrons. 
In these systems, where M = Zn, Cd, or Hg, the low-energy electronic structure is primarily composed of $s$ and $p$ 
states, which are typically regarded as weakly correlated due to their spatially extended nature and reduced Coulomb 
repulsion. Nonetheless, several compounds in this class exhibit semimetallic or even topologically nontrivial features 
that make their screening properties particularly intriguing. Our cRPA calculations reveal that, despite their weaker 
on-site interactions, these systems often exhibit non-negligible long-range Coulomb interactions, especially in the 
$\beta$-phase structures. For example, in $\beta_2$-CdGa$_2$Se$_4$, the sixth-neighbor interaction reaches 0.26~eV, 
exceeding that of some semiconducting $\alpha$-phase systems with $d$ orbitals (see Table~S3 in the Supplementary 
Material).

To provide a comprehensive reference for future studies, we include in Table~S2 of the Supplementary Information 
the effective (partially screened) Coulomb parameters \( U \), \( U' \), and \( J \), as well as their fully screened 
counterparts \( \tilde{U} \), \( \tilde{U}' \), and \( \tilde{J} \), evaluated for multiple correlated subspaces in the 
32-valence-electron systems. In Table~S3, we further present distance-resolved off-site Coulomb interactions \( U(r) \) 
up to the sixth nearest neighbor. These data confirm that long-range interactions in $sp$-electron semimetals can be 
significant and underscore the potential relevance of nonlocal correlation effects in these systems.

\subsection*{Coulomb interactions in 33-valence electron compounds}

\begin{table*}[tp]
\caption{Crystal symmetry, lattice parameter, ground state, correlated subspace, bare Coulomb interaction $V$, 
partially screened Hubbard-Kanamori parameters [$U$, $U^{\prime}$, $J$], fully screened counterparts [$\tilde{U}$, 
$\tilde{U}^{\prime}$, $\tilde{J}$] (in eV) for $d$ orbitals of TMs in MA$_2$Z$_4$ (M=V, Nb, Ta; A=Si, Ge; Z=N, P, As) compounds.
Lattice parameters are taken from Ref.~\onlinecite{Wang}. Abbreviations: SGS — spin gapless semiconductor; SGM — spin gapped metal.}
	\begin{ruledtabular}
 	\begin{tabular}{lccccccccccc}
		\multirow{2}{*}{MA$_2$Z$_4$} & \multirow{2}{*}{Phase} & a & Ground & Correlated & $V$ & $U$ & $U^{\prime}$ & $J$ & $\tilde{U}$ & $\tilde{U}^{\prime}$ & $\tilde{J}$ \\
 		&  & (\AA) & State & Subspace & (eV) & (eV) & (eV) & (eV) & (eV) & (eV) & (eV) \\ \hline
 		\multirow{3}{*}{VSi$_2$N$_4$} & \multirow{3}{*}{$\alpha_1$} & \multirow{3}{*}{2.88} & \multirow{3}{*}{\begin{tabular}[c]{@{}c@{}} SGS\\ (type-II)\end{tabular}} & $d_{z^2}$ & 7.06 & 1.48 & - & - & 0.40 & - & - \\
 		&  &  &  & $d_{z^2}$+$d_{xy}$+$d_{x^2-y^2}$ & 14.79 & 2.66 & 1.81 & 0.46 & 1.03 & 0.45 & 0.29 \\
 		&  &  &  & d & 13.76 & 2.40 & 1.84 & 0.37 & 1.07 & 0.51 & 0.27 \\
 		\multirow{3}{*}{VSi$_2$P$_4$} & \multirow{3}{*}{$\delta_4$} & \multirow{3}{*}{3.48} & \multirow{3}{*}{\begin{tabular}[c]{@{}c@{}} SGS\\ (type-II)\end{tabular}} & $d_{z^2}$ & 5.68 & 0.80 & - & - & 0.22 & - & - \\
 		&  &  &  & $d_{z^2}$+$d_{xy}$+$d_{x^2-y^2}$ & 13.29 & 1.44 & 0.75 & 0.36 & 0.79 & 0.31 & 0.25 \\
 		&  &  &  & d & 12.15 & 1.26 & 0.77 & 0.28 & 0.76 & 0.34 & 0.21 \\
 		\multirow{3}{*}{VSi$_2$As$_4$} & \multirow{3}{*}{$\alpha_2$} & \multirow{3}{*}{3.64} & \multirow{3}{*}{\begin{tabular}[c]{@{}c@{}} SGS\\ (type-I)\end{tabular}} & $d_{z^2}$ & 5.79 & 0.73 & - & - & 0.25 & - & - \\
 		&  &  &  & $d_{z^2}$+$d_{xy}$+$\textnormal d_{x^2-y^2}$ & 13.60 & 1.38 & 0.67 & 0.37 & 0.72 & 0.25 & 0.24 \\
 		&  &  &  & d & 12.30 & 1.23 & 0.73 & 0.28 & 0.73 & 0.29 & 0.21 \\
 		\multirow{3}{*}{VGe$_2$N$_4$} & \multirow{3}{*}{$\beta_2$} & \multirow{3}{*}{3.05} & \multirow{3}{*}{\begin{tabular}[c]{@{}c@{}} SGM\\ (pn-type)\end{tabular}} & $d_{z^2}$ & 6.35 & 0.43 & - & - & 0.17 & - & - \\
 		&  &  &  & $d_{z^2}+\textnormal e_{g}$ & 14.48 & 2.63 & 1.86 & 0.39 & 0.64 & 0.16 & 0.22 \\
 		&  &  &  & d & 14.38 & 2.15 & 1.51 & 0.37 & 0.80 & 0.21 & 0.25 \\
 		\multirow{3}{*}{VGe$_2$P$_4$} & \multirow{3}{*}{$\alpha_2$} & \multirow{3}{*}{3.56} & \multirow{3}{*}{\begin{tabular}[c]{@{}c@{}}SGM\\ (pn-type)\end{tabular}} & $d_{z^2}$ & 5.84 & 0.78 & - & - & 0.28 & - & - \\
 		&  &  &  & $d_{z^2}$+$d_{xy}$+$d_{x^2-y^2}$ & 13.75 & 1.50 & 0.77 & 0.37 & 0.81 & 0.30 & 0.26 \\
 		&  &  &  & d & 12.42 & 1.32 & 0.81 & 0.29 & 0.80 & 0.35 & 0.23 \\
 		\multirow{3}{*}{VGe$_2$As$_4$} & \multirow{3}{*}{$\alpha_2$} & \multirow{3}{*}{3.72} & \multirow{3}{*}{\begin{tabular}[c]{@{}c@{}} SSM \end{tabular}} & $d_{z^2}$ & 5.77 & 0.66 & - & - & 0.24 & - & - \\
 		&  &  &  & $d_{z^2}$+$d_{xy}$+$d_{x^2-y^2}$ & 13.80 & 1.37 & 0.65 & 0.37 & 0.71 & 0.23 & 0.24 \\
 		&  &  &  & d & 12.33 & 1.20 & 0.69 & 0.28 & 0.71 & 0.28 & 0.21 \\
 		\multirow{3}{*}{NbSi$_2$N$_4$} & \multirow{3}{*}{$\alpha_1$} & \multirow{3}{*}{2.96} & \multirow{3}{*}{\begin{tabular}[c]{@{}c@{}} SGM\\ (pn-type)\end{tabular}} & $d_{z^2}$ & 5.52 & 1.51 & - & - & 0.38 & - & - \\
 		&  &  &  & $d_{z^2}$+$d_{xy}$+$d_{x^2-y^2}$ & 10.15 & 2.16 & 1.62 & 0.31 & 1.01 & 0.55 & 0.24 \\
 		&  &  &  & d & 9.76 & 2.08 & 1.70 & 0.26 & 1.05 & 0.61 & 0.22 \\
        \multirow{3}{*}{NbSi$_2$P$_4$} & \multirow{3}{*}{$\alpha_1$} & \multirow{3}{*}{3.53} & \multirow{3}{*}{\begin{tabular}[c]{@{}c@{}}Cold-Metal\\ (p-type)\end{tabular}} & $d_{z^2}$ & 4.02 & 0.71 & - & - & 0.21 & - & - \\
 		&  &  &  & $d_{z^2}$+$d_{xy}$+$d_{x^2-y^2}$ & 8.56 & 1.12 & 0.70 & 0.23 & 0.75 & 0.38 & 0.19 \\
 		&  &  &  & d & 8.06 & 1.06 & 0.74 & 0.19 & 0.74 & 0.41 & 0.17 \\
 		\multirow{3}{*}{NbSi$_2$As$_4$} & \multirow{3}{*}{$\alpha_2$} & \multirow{3}{*}{3.68} & \multirow{3}{*}{\begin{tabular}[c]{@{}c@{}}Cold-Metal\\ (p-type)\end{tabular}} & $d_{z^2}$ & 3.95 & 0.63 & - & - & 0.22 & - & - \\
 		&  &  &  & $d_{z^2}$+$d_{xy}$+$d_{x^2-y^2}$ & 9.50 & 1.15 & 0.65 & 0.27 & 0.66 & 0.33 & 0.18 \\
 		&  &  &  & d & 8.30 & 0.96 & 0.62 & 0.19 & 0.59 & 0.37 & 0.11 \\
 	    \multirow{3}{*}{NbGe$_2$N$_4$} & \multirow{3}{*}{$\alpha_1$} & \multirow{3}{*}{3.09} & \multirow{3}{*}{\begin{tabular}[c]{@{}c@{}} SGM\\ (pn-type)\end{tabular}} & $d_{z^2}$ & 5.44 & 1.38 & - & - & 0.33 & - & - \\
 		&  &  &  & $d_{z^2}$+$d_{xy}$+$d_{x^2-y^2}$ & 9.68 & 1.95 & 1.43 & 0.29 & 0.87 & 0.45 & 0.21 \\
 		&  &  &  & d & 8.10 & 1.61 & 1.33 & 0.22 & 0.77 & 0.45 & 0.18 \\
 		\multirow{3}{*}{NbGe$_2$P$_4$} & \multirow{3}{*}{$\alpha_2$} & \multirow{3}{*}{3.62} & \multirow{3}{*}{\begin{tabular}[c]{@{}c@{}}Cold-Metal\\ (p-type)\end{tabular}} & $d_{z^2}$ & 4.06 & 0.72 & - & - & 0.25 & - & - \\
 		&  &  &  & $d_{z^2}$+$d_{xy}$+$ d_{x^2-y^2}$ & 8.98 & 1.19 & 0.72 & 0.24 & 0.67 & 0.37 & 0.16 \\
 		&  &  &  & d & 8.36 & 1.09 & 0.76 & 0.20 & 0.64 & 0.43 & 0.11 \\
 		\multirow{3}{*}{NbGe$_2$As$_4$} & \multirow{3}{*}{$\alpha_2$} & \multirow{3}{*}{3.77} & \multirow{3}{*}{\begin{tabular}[c]{@{}c@{}}Cold-Metal\\ (p-type)\end{tabular}} & $d_{z^2}$ & 4.00 & 0.61 & - & - & 0.22 & - & - \\
 		&  &  &  & $d_{z^2}$+$d_{xy}$+$ d_{x^2-y^2}$ & 9.26 & 1.12 & 0.63 & 0.25 & 0.72 & 0.31 & 0.21 \\
 		&  &  &  & d & 8.63 & 0.92 & 0.70 & 0.13 & 0.61 & 0.38 & 0.12 \\        
 		\multirow{3}{*}{TaSi$_2$N$_4$} & \multirow{3}{*}{$\alpha_1$} & \multirow{3}{*}{2.97} & \multirow{3}{*}{\begin{tabular}[c]{@{}c@{}}Cold-Metal\\ (pn-type)\end{tabular}} & $d_{z^2}$ & 5.36 & 1.57 & - & - & 0.39 & - & - \\
 		&  &  &  & $d_{z^2}$+$d_{xy}$+$d_{x^2-y^2}$ & 9.35 & 2.14 & 1.63 & 0.30 & 1.03 & 0.59 & 0.24 \\
 		&  &  &  & d & 8.58 & 1.93 & 1.62 & 0.24 & 1.00 & 0.63 & 0.21 \\
 		\multirow{3}{*}{TaSi$_2$P$_4$} & \multirow{3}{*}{$\alpha_1$} & \multirow{3}{*}{3.54} & \multirow{3}{*}{\begin{tabular}[c]{@{}c@{}}Cold-Metal\\ (p-type)\end{tabular}} & $d_{z^2}$ & 4.31 & 0.81 & - & - & 0.29 & - & - \\
 		&  &  &  & $d_{z^2}$+$d_{xy}$+$d_{x^2-y^2}$ & 8.00 & 1.12 & 0.72 & 0.22 & 0.75 & 0.39 & 0.19 \\
 		&  &  &  & d & 7.60 & 1.05 & 0.75 & 0.18 & 0.74 & 0.42 & 0.16 \\
 		\multirow{3}{*}{TaSi$_2$As$_4$} & \multirow{3}{*}{$\alpha_2$} & \multirow{3}{*}{3.68} & \multirow{3}{*}{\begin{tabular}[c]{@{}c@{}}Cold-Metal\\ (p-type)\end{tabular}} & $d_{z^2}$ & 3.90 & 0.67 & - & - & 0.23 & - & - \\
 		&  &  &  & $d_{z^2}$+$d_{xy}$+$d_{x^2-y^2}$ & 8.70 & 1.14 & 0.68 & 0.26 & 0.67 & 0.36 & 0.17 \\
 		&  &  &  & d & 7.71 & 0.86 & 0.66 & 0.12 & 0.60 & 0.40 & 0.10 \\     
 		\multirow{3}{*}{TaGe$_2$N$_4$} & \multirow{3}{*}{$\alpha_1$} & \multirow{3}{*}{3.08} & \multirow{3}{*}{\begin{tabular}[c]{@{}c@{}} SGM\\ (pn-type)\end{tabular}} & $d_{z^2}$ & 5.17 & 1.42 & - & - & 0.35 & - & - \\
 		&  &  &  & $d_{z^2}$+$d_{xy}$+$\textnormal d_{x^2-y^2}$ & 8.01 & 1.70 & 1.28 & 0.23 & 0.80 & 0.45 & 0.18 \\
 		&  &  &  & d & 8.39 & 1.78 & 1.45 & 0.23 & 0.91 & 0.54 & 0.20 \\
 		\multirow{3}{*}{TaGe$_2$P$_4$} & \multirow{3}{*}{$\alpha_2$} & \multirow{3}{*}{3.61} & \multirow{3}{*}{\begin{tabular}[c]{@{}c@{}}Cold-Metal\\ (p-type)\end{tabular}} & $d_{z^2}$ & 4.41 & 0.82 & - & - & 0.28 & - & - \\
 		&  &  &  & $d_{z^2}$+$d_{xy}$+$d_{x^2-y^2}$ & 8.30 & 1.15 & 0.72 & 0.22 & 0.68 & 0.39 & 0.15 \\
 		&  &  &  & d & 7.64 & 1.05 & 0.74 & 0.18 & 0.64 & 0.43 & 0.10 \\
 		\multirow{3}{*}{TaGe$_2$As$_4$} & \multirow{3}{*}{$\alpha_2$} & \multirow{3}{*}{3.77} & \multirow{3}{*}{\begin{tabular}[c]{@{}c@{}}Cold-Metal\\ (p-type)\end{tabular}} & $d_{z^2}$ & 4.32 & 0.70 & - & - & 0.24 & - & - \\
 		&  &  &  & $d_{z^2}$+$d_{xy}$+$d_{x^2-y^2}$ & 8.56 & 1.11 & 0.64 & 0.24 & 0.73 & 0.33 & 0.20 \\
 		&  &  &  & d & 8.02 & 1.03 & 0.68 & 0.20 & 0.62 & 0.40 & 0.11 
	\end{tabular}
\label{table2}
\end{ruledtabular}
\end{table*}

We now turn to the class of MA$_2$Z$_4$ compounds with 33 valence electrons, where M = V, Nb, or Ta; 
A = Si or Ge; and Z = N, P, or As. These materials exhibit a diverse range of electronic ground states, 
including spin-gapless semiconductors (SGSs), spin-gapped metals (SGMs), and cold metals (see Fig.~S4 in
the Supplementary Information) characterized by a narrow half-filled band crossing the Fermi 
level \cite{latychevskaia2025mosi2n4,tho2023ma2z4,wang2008proposal,csacsiouglu2025spin,zhang2024high}. 
This narrow-band feature is often indicative of enhanced electronic correlations, although a detailed 
analysis of correlation strength in terms of the ratio $U/W$ will be presented in a separate subsection. 
A notable distinction arises in their magnetic character: V-based systems tend to favor magnetically 
ordered ground states and frequently realize SGS or SGM phases, whereas Nb- and Ta-based compounds are 
predominantly nonmagnetic and typically exhibit p-type or pn-type cold metallic behavior~\cite{Wang,Feng-2022,Feng}. 
A few exceptions exist, notably $\alpha_1$-NbSi$_2$N$_4$, $\alpha_1$-NbGe$_2$N$_4$, and 
$\alpha_1$-TaGe$_2$N$_4$, which display small exchange splittings and feature well-defined gaps above 
and below the Fermi level (see Fig.~S5 in the Supplementary Information). These systems are classified 
as SGMs, or equivalently, magnetic cold metals.

To determine the Coulomb interactions in these systems, we perform cRPA and full RPA calculations using a 
non-magnetic reference state for all compounds, considering three correlated subspaces: the $d_{z^2}$ 
orbital, a three-orbital set ($d_{z^2}+d_{xy}+d_{x^2-y^2}$), and the full $d$-manifold. The resulting
interaction parameters are summarized in Table~\ref{table2}, which also indicates the magnetic ground state
of each compound for completeness. This consistent dataset provides a basis for comparing with the 
34-valence-electron systems discussed earlier and enables future investigations of correlation effects 
across the MA$_2$Z$_4$ family.

Compared to the 34-valence-electron compounds, the bare Coulomb interactions $V$ in the 33-electron systems 
are systematically reduced by 1--2 eV. This reduction stems from the smaller nuclear charge of V, Nb, and 
Ta relative to Cr, Mo, and W, which leads to less contracted atomic orbitals, more spatially extended Wannier 
functions, and consequently smaller Coulomb interaction parameters. These trends align with prior observations
in related materials~\cite{Ramezani,Karbalaee_Aghaee_2022,yekta_2021} and reflect the chemical effects of 
reduced nuclear charge and orbital delocalization in lighter transition metals.

A similar trend is observed in the partially screened interactions $U$, $U'$, and $J$, which decrease from 
Cr/Mo/W-based compounds to their V/Nb/Ta counterparts. This behavior arises from the combined influence 
of reduced orbital localization and enhanced screening. For example, in the cold metallic compound 
$\alpha_1$-NbSi$_2$N$_4$, the intra-orbital $U$ is 2.16 eV, while the fully screened $\tilde{U}$ drops 
to 1.01 eV —less than half of $U$— owing to strong screening from low-energy $d \rightarrow d$ transitions 
within the narrow $d_{z^2}$ band crossing the Fermi level.

This pronounced reduction of $\tilde{U}$ relative to $U$ highlights the critical role of low-energy screening
channels in metallic systems and motivates a real-space analysis of the distance dependence of Coulomb interactions.
To this end, we analyze the spatial decay of $U(r)$ and $\tilde{U}(r)$ in selected 33-valence-electron 
MA$_2$Z$_4$ compounds. For all materials in this class, $U(r)$ is computed up to the sixth nearest neighbor, 
with the full dataset provided in Table~S4 of the Supplementary Information. To illustrate representative 
screening behavior, Fig.~\ref{fig:long-MoW} (panels e--h) shows the real-space profiles of $V(r)$, $U(r)$, 
and $\tilde{U}(r)$ for the $\alpha_1$ and $\beta_2$ phases of NbGe$_2$N$_4$ and TaSi$_2$N$_4$.

In the $\alpha_1$ phase, both compounds posses a single $d_{z^2}$-derived band crossing the 
Fermi level. Accordingly, we restrict the correlated subspace to the $d_{z^2}$ orbital. In this case, the 
partially screened interaction $U(r)$ remains long-ranged, extending well beyond twelve lattice constants. 
By contrast, the fully screened $\tilde{U}(r)$ decays more rapidly and becomes negligible after approximately 
six neighbors, reflecting efficient screening by low-energy metallic states near the Fermi level.

\begin{table}[tp]
\caption{The on-site ($U^{\mathrm{eff}}_{00}$) and the off-site ($U^{\mathrm{eff}}_{01}$) effective Coulomb 
interactions are calculated for a specific atomic rearrangement pattern within the $\beta_{2}$ phases of MA$_2$Z$_4$ 
structure, known as the star-of-David configuration.}
\begin{ruledtabular}
\begin{tabular}{lcccc}
        MA$_2$Z$_4$ & Phase & Correlated Subspace & $U_{00}$(eV) & $U_{01}$(eV) \\ \hline
	\rule{0pt}{4mm}%
	\multirow{2}{*}{VGe$_2$N$_4$} & \multirow{2}{*}{$\beta_{2}$} & $d_{z^2}$ & 0.17 & 0.08 \\
	&  & $d_{z^2}+ e_g$ & 1.02 & 0.63 \\
	\multirow{2}{*}{NbSi$_2$N$_4$} & \multirow{2}{*}{$\beta_{2}$} & $d_{z^2}$ & 0.94 & 0.72 \\
	&  & $d_{z^2}+e_g$ & 0.54 & 0.19 \\
	\multirow{2}{*}{NbGe$_2$N$_4$} & \multirow{2}{*}{$\beta_{2}$} & $d_{z^2}$ & 0.91 & 0.65 \\
	&  & $d_{z^2}+e_g$ & 0.52 & 0.19 \\
	\multirow{2}{*}{NbSi$_2$P$_4$} & \multirow{2}{*}{$\beta_{2}$} & $d_{z^2}$ & 0.16 & 0.09 \\
	&  & $d_{z^2}+e_g$ & 0.21 & 0.09 \\
	\multirow{2}{*}{TaSi$_2$N$_4$} & \multirow{2}{*}{$\beta_{2}$} & $d_{z^2}$ & 1.08 & 0.78 \\
	&  & $d_{z^2}+e_g$ & 0.61 & 0.22 \\
	\multirow{2}{*}{TaGe$_2$N$_4$} & \multirow{2}{*}{$\beta_{2}$} & $d_{z^2}$ & 0.95 & 0.68 \\
	&  & $d_{z^2}+e_g$ & 0.51 & 0.18 \\
	\multirow{2}{*}{TaSi$_2$P$_4$} & \multirow{2}{*}{$\beta_{2}$} & $d_{z^2}$ & 0.16 & 0.09 \\
	&  & $d_{z^2}+ e_g$ & 0.21 & 0.09
    \end{tabular}
     \label{table-x}
    \end{ruledtabular}
    \end{table}

In the $\beta_2$ phase, which is metastable but frequently emerges as the second-lowest-energy structure, both 
NbGe$_2$N$_4$ and TaSi$_2$N$_4$ remain metallic, exhibiting n-type cold-metal behavior (see Ref.\,\onlinecite{Wang} 
and Table~S5 in the Supplementary Information). Unlike the $\alpha_1$ phase, however, the low-energy states in this 
configuration involve more entangled orbital characters, necessitating a three-orbital correlated subspace 
($d_{z^2} + e_g$) for accurate modeling. Within this subspace, both the partially screened interaction $U(r)$ and 
the fully screened interaction $\tilde{U}(r)$ decay rapidly with distance, vanishing beyond six lattice spacings 
(see Fig.~\ref{fig:long-MoW}). This short-range behavior reflects more efficient metallic screening in the $\beta_2$ phase. 
The corresponding real-space Coulomb interaction parameters, both on-site and off-site, are provided in Table~S5 and Table~S6 
of the Supplementary Information. These values are particularly relevant for understanding correlation effects in 
structurally reconstructed $\beta_2$ phases, where charge-density-wave (CDW) instabilities are known to emerge.

To connect with these CDW phenomena, we analyze the effective Coulomb interaction parameters relevant for the 
star-of-David (SOD) reconstruction [see Fig.~\ref{fig:struc}(f)], which has been theoretically predicted to occur
at low temperatures in $\beta_2$-phase MA$_2$Z$_4$ compounds~\cite{Wang,Nano-2024-wang}. While our calculations are
performed for the undistorted $\beta_2$ lattice, they provide critical input for modeling the insulating SOD 
phase. Using the methodology described in Eq.~\ref{U_eff_SOD}, we extract both the on-site ($U^{\mathrm{eff}}_{00}$) 
and nearest-neighbor ($U^{\mathrm{eff}}_{01}$) screened Coulomb interactions for several systems with Z = N, as 
listed in Table~\ref{table-x}. We find $U^{\mathrm{eff}}_{00}$ values in the range of 0.9–1.1 eV, significantly 
higher than the 0.65–0.80 eV reported for prototypical CDW systems like NbS$_2$ and TaS$2$~\cite{Ramezani,Kim-2023}. 
In addition, the off-site interaction $U^{\mathrm{eff}}_{01}$ reaches up to 75\% of the on-site value, indicating 
strong nonlocal interactions. These features, combined with the narrow bandwidth of the reconstructed SOD phase 
($W \approx 0.02$ eV)~\cite{Kim-2023,Nano-2024-wang}, point toward a strong-coupling regime that favors correlation-driven CDW 
order. The extracted parameters thus form a reliable foundation for future many-body simulations based on extended 
Hubbard models or methods such as DFT+$U$ and DFT+DMFT. Having established the importance of nonlocal interactions 
in charge-ordered systems, we now shift our attention to Mn-based MA$_2$Z$_4$ compounds, where strong correlations 
give rise to ferromagnetism and topological phases.

\begin{table*}[tp]
\caption{Lattice parameters, electronic ground states, correlated subspaces, and Coulomb interaction 
parameters for the $d$ orbitals of transition-metal atoms in  two-dimensional MnA$_2$Z$_4$ 
compounds (A = Bi, Sb; Z = Se, Te) with 41 valence electrons. The listed quantities include the bare 
intra-orbital Coulomb interaction $V$, the partially screened Hubbard–Kanamori parameters [$U$, $U^{\prime}$, $J$], 
and their fully screened counterparts [$\tilde{U}$, $\tilde{U}'$, $\tilde{J}$]. For each compound, 
values are reported for two distinct correlated subspaces: 
a three-orbital model ($d_{z^2}$ + $e_g$) and the full $d$-orbital manifold. 
Lattice constants are taken from Ref.~\onlinecite{Zhu2021,Wimmer2021}. 
The abbreviation FM-SC denotes a ferromagnetic semiconductor.}
\begin{ruledtabular}
\begin{tabular}{lccccccccccc}
\multirow{2}{*}{MA$_2$Z$_4$} & \multirow{2}{*}{Phase} & a & Ground & Correlated &  $V$  & $U$ & $U^{\prime}$ & $J$  & $\tilde{U}$ & $\tilde{U}^{\prime}$ & $\tilde{J}$ \\
&  & (\AA) & State & Subspace & (eV) & (eV) & (eV) & (eV) & (eV) & (eV) & (eV) \\ \hline
\multirow{2}{*}{MnSb$_2$Se$_4$} & \multirow{2}{*}{$\beta_{5}$} & \multirow{2}{*}{3.87} & \multirow{2}{*}{FM-SC} & $d_{z^2}+e_{g}$ & 12.01 & 0.45 & 0.16 & 0.14 & 0.27 & 0.14 & 0.06 \\
&  &  &  & $d$ & 11.52 & 1.29 & 0.89 & 0.26 & 0.30 & 0.15 & 0.08 \\
\multirow{2}{*}{MnSb$_2$Te$_4$} & \multirow{2}{*}{$\beta_{5}$} & \multirow{2}{*}{4.23} & \multirow{2}{*}{FM-SC} & $d_{z^2}+e_{g}$ & 14.16 & 0.90 & 0.40 & 0.23 & 0.44 & 0.21 & 0.11 \\
&  &  &  & $d$ & 10.45 & 1.04 & 0.70 & 0.22 & 0.39 & 0.22 & 0.10 \\
\multirow{2}{*}{MnBi$_2$Se$_4$} & \multirow{2}{*}{$\beta_{5}$} & \multirow{2}{*}{3.93} & \multirow{2}{*}{FM-SC} & $d_{z^2}+e_{g}$ & 14.47 & 0.81 & 0.39 & 0.22 & 0.66 & 0.52 & 0.07 \\
&  &  &  & $d$ & 12.77 & 1.67 & 1.15 & 0.33 & 0.73 & 0.53 & 0.09 \\
\multirow{2}{*}{MnBi$_2$Te$_4$} & \multirow{2}{*}{$\beta_{5}$} & \multirow{2}{*}{4.34} & \multirow{2}{*}{FM-SC} & $d_{z^2}+e_{g}$ & 14.12 & 1.07 & 0.50 & 0.29 & 0.29 & 0.18 & 0.10 \\
&  &  &  & $d$ & 11.56 & 1.21 & 0.80 & 0.26 & 0.40 & 0.20 & 0.10 
\end{tabular}
\label{table2x}
\end{ruledtabular}
\end{table*}


\subsection*{Coulomb interactions in 41-valence-electron Mn-based topological materials}

We now examine the Coulomb interaction parameters in a class of Mn-based MA$_2$Z$_4$ compounds 
with 41 valence electrons, where A = Bi or Sb and Z = Se or Te. These materials, which all 
crystallize in the $\beta_5$ phase, exhibit robust ferromagnetic semiconducting (FM-SC) ground 
states and are widely recognized for their topologically nontrivial properties~\cite{Zhu2021,Guo2024,Qiu2025-axion,Vyazovskaya2025,Otrokov2019,Gao2024_AFMdiode,Zhang2025_edgeSupercurrent,Wimmer2021}. Among them, 
$\beta_5$-MnBi$_2$Te$_4$ has emerged as a prototypical intrinsic magnetic topological insulator, 
exhibiting hallmark phenomena such as the quantum anomalous Hall effect, axion electrodynamics, 
and high-Chern-number states in thin-film form~\cite{Guo2024,Qiu2025-axion,Vyazovskaya2025,Otrokov2019}. These magnetic and topological
features originate from the interplay between the localized Mn $d$ orbitals and the surrounding
Bi/Te $p$ states, motivating a detailed analysis of the effective Coulomb interaction parameters.

To analyze interaction trends in these Mn-based topological compounds, we evaluate the bare 
Coulomb interaction $V$, the partially screened Hubbard–Kanamori parameters ($U$, $U'$, $J$), 
and their fully screened counterparts ($\tilde{U}$, $\tilde{U}'$, $\tilde{J}$) for four 
representative systems: $\beta_5$-MnBi$_2$Te$_4$, $\beta_5$-MnBi$_2$Se$_4$, $\beta_5$-MnSb$_2$Te$_4$, 
and $\beta_5$-MnSb$_2$Se$_4$. These parameters, obtained from cRPA (RPA) calculations in the 
non-magnetic reference state, are summarized in Table~\ref{table2x}. For each material, 
we consider two correlated subspaces: a three-orbital model comprising $d_{z^2}$ and $e_g$ 
orbitals, and the full five-orbital $d$-manifold.

Several trends emerge from the data in Table~\ref{table2x}. Most notably, the bare Coulomb 
interaction $V$ is consistently larger in the three-orbital subspace, reaching up to 14.5\,eV 
in MnBi$_2$Se$_4$. This behavior reflects the more localized nature of the $d_{z^2}$ and $e_g$ 
orbitals that dominate the electronic states near the Fermi level. As shown in Fig.~\ref{fig:interpol}(e), 
the remaining two Mn $d$ orbitals lie higher in energy and exhibit significant $p$–$d$ hybridization, 
leading to greater spatial delocalization and thus their inclusion in the full five-orbital $d$-manifold 
reduces the averaged bare interaction. Upon inclusion of screening, the partially screened $U$ 
values are substantially reduced, typically falling in the range of 0.45–1.67\,eV depending on the 
compound and subspace. The exchange parameter $J$ ranges from 0.14 to 0.33\,eV, indicating moderate 
intra-atomic exchange interactions that are relevant for stabilizing the ferromagnetic ground state
in these materials. To complement these local interaction values, the spatial decay of the partially 
screened effective \( U \) values—extending up to the sixth nearest neighbor—is presented in the Supplementary 
Information (Table S7), further illustrating the extended nature of Coulomb interactions in the $\beta_5$ phase.

Across the series, a clear chemical trend is observed. Substituting Bi with Sb generally lowers 
both the bare and screened interaction parameters, consistent with the smaller atomic radius and 
higher electronegativity of Sb. These factors lead to shorter Mn–A bond distances and enhanced 
hybridization, which in turn increase the overall interaction strength. Similarly, replacing Te 
with the more localized Se results in slightly larger $V$ values, as expected. The partially 
screened Hubbard $U$ values for the full $d$-orbital manifold are consistently larger than those 
for the three-orbital model. This reflects the fact that, in the full subspace, more screening 
channels are excluded from the polarization function, leading to reduced screening and hence larger 
interaction values. In contrast, the three-orbital model retains screening contributions from the 
remaining $d$ states, resulting in lower $U$ values. The fully screened interactions $\tilde{U}$ 
and $\tilde{U}'$, though systematically smaller than their partially screened counterparts, remain 
in the range of 0.27–0.73\,eV.

These Coulomb interaction parameters provide a quantitative basis for modeling Mn-based magnetic 
topological insulators using extended Hubbard models, DFT+DMFT,  or DFT+$U$($+V$) approaches. They 
also offer key insights into how chemical substitution and orbital selection influence the balance 
between electronic localization and screening, an interplay that underpins the emergence of magnetic 
and topological phases in this material family.

More broadly, the results for 32-, 33-, 34-, and 41-valence-electron MA$_2$Z$_4$ monolayers reveal
systematic trends in both the strength and spatial extent of Coulomb interactions. These trends reflect 
a complex interplay between orbital localization, screening efficiency, and structural phase. For instance, 
semiconducting 34-valence-electron compounds exhibit relatively strong yet spatially extended interactions, 
while cold metals in the 33-valence-electron class show sharp contrasts between partially and fully screened 
interactions, particularly in phases with narrow bands crossing the Fermi level. In metastable $\beta_2$ 
structures, enhanced metallic screening significantly suppresses the long-range components of the interaction. 
Overall, these findings provide a foundation for describing correlation effects across this material family
and offer valuable context for interpreting the collective excitations and magnetic behavior discussed in 
the subsequent sections.

\subsection*{Correlation strength and magnetic instabilities in  MA$_2$Z$_4$ materials}

To assess the degree of electronic correlations across the MA\(_2\)Z\(_4\) family, we evaluate 
the ratio of the partially screened on-site Coulomb interaction \( U \) to the corresponding 
bandwidth \( W \), commonly denoted as \( U/W \). This dimensionless parameter serves as a key
indicator of correlation strength, distinguishing weakly correlated metals from strongly correlated 
systems, and offers insight into the propensity for magnetic ordering or Mott-like behavior. 
Typically, systems with \( U/W < 1 \) are considered weakly correlated, those with \( U/W \approx 1 \) 
moderately correlated, and those with \( U/W > 1 \) strongly correlated. Given the diversity of 
orbital characters near the Fermi level, we compute \( U/W \) for three distinct correlated subspaces: 
a single-orbital model (typically dominated by the most localized \( d_{z^2} \) orbital), 
a three-orbital model (\( d_{z^2} + e_g \)), and the full five-orbital \( d \)-manifold.

Among the 33-valence-electron compounds, characterized by a half-filled band at the Fermi 
level, we find large values of the correlation strength \( U/W \). In the single-orbital 
subspace, \( U/W \) frequently exceeds unity. These high ratios reflect narrow bandwidths and 
reduced metallic screening, resulting from the presence of both internal and external energy 
gaps around \( E_F \). While the three- and five-orbital subspaces yield lower \( U/W \) 
values due to increased hybridization and broader band dispersions, signatures of strong 
coupling persist. This trend points to an inherent tendency toward magnetic ordering, 
consistent with our spin-polarized DFT results and prior theoretical studies.

\begin{table*}[tp]
\caption{Valence electron count (VEC), correlation strength \( U/W \), Stoner parameter \( I \) (in eV), 
non-magnetic density of states at the Fermi level \( N(E_F) \) (in states/eV), Stoner criterion \( I N(E_F) \), 
and spin-polarized magnetic moment \( m \) (in \( \mu_B \)) for selected MA\(_2\)Z\(_4\) compounds with 
33, 34, and 41 valence electrons. Values are reported for three correlated subspaces: the single-orbital 
model (predominantly \( d_{z^2} \)), the three-orbital model (\( d_{z^2} + e_g \)), and the full five-orbital
\( d \)-manifold. The table enables a comparative assessment of correlation strength and the tendency toward
ferromagnetic ordering, as indicated by the Stoner criterion \( I N(E_F) > 1 \), and includes the resulting 
magnetic moments from spin-polarized DFT calculations.}
	\begin{ruledtabular}
	\begin{tabular}{lcccccccccccccccc}
		& & \multicolumn{4}{c}{1-orbital subspace} &  & \multicolumn{4}{c}{3-orbital subspace} &  & \multicolumn{4}{c}{5-orbital subspace} &  \\ \cline{3-6} \cline{8-11} \cline{13-16}
		MA$_2$Z$_4$ & VEC & $U/W$ & $I$ & $N(E_{\textnormal F})$ & $I.N(E_{\textnormal F})$ &  & $U/W$ & $I$ & $N(E_{\textnormal F})$ & $I.N(E_{\textnormal F})$ &  & $U/W$ & $I$ & $N(E_{\textnormal F})$ & $I.N(E_{\textnormal F})$ & $m_{\mathrm{M}}$ \\      
        \hline
			\rule{0pt}{3mm}%
		$\alpha_{1}$-VSi$_2$N$_4$ &33 & 1.37 & 1.48 & 1.31 & 1.94 &  & 0.80 & 1.50 & 2.53 & 3.80 &  & 0.54 & 0.97 & 2.53 & 2.45 & 1.03 \\
		$\delta_{4}$-VSi$_2$P$_4$ &33& 1.14 & 0.80 & 1.42 & 1.14 &  & 0.48 & 0.95 & 2.67 & 2.54 &  & 0.38 & 0.61 & 2.67 & 1.63 & 1.06 \\
		$\alpha_{2}$-VSi$_2$As$_4$ &33& 1.03 & 0.73 & 1.66 & 1.21 &  & 0.48 & 0.94 & 3.27 & 3.07 &  & 0.38 & 0.60 & 3.27 & 1.96 & 1.06 \\
		$\beta_{2}$-VGe$_2$N$_4$ &33& 0.54 & 0.43 & 1.40 & 0.60 &  & 1.62 & 1.38 & 2.15 & 2.97 &  & 0.57 & 0.90 & 3.06 & 2.75 & 1.12 \\
		$\alpha_{2}$-VGe$_2$P$_4$ &33& 0.88 & 0.78 & 1.37 & 1.07 &  & 0.47 & 0.98 & 2.75 & 2.70 &  & 0.38 & 0.63 & 2.75 & 1.73 & 0.79 \\
		$\alpha_{2}$-VGe$_2$As$_4$ &33& 0.86 & 0.66 & 1.62 & 1.07 &  & 0.50 & 0.94 & 3.29 & 3.09 &  & 0.37 & 0.60 & 3.29 & 1.97 & 1.06 \\
		$\alpha_{1}$-NbSi$_2$N$_4$ &33& 1.24 & 1.51 & 1.03 & 1.56 &  & 0.39 & 1.14 & 1.90 & 2.17 &  & 0.33 & 0.76 & 1.90 & 1.44 & 0.43 \\
		$\alpha_{1}$-NbSi$_2$P$_4$ &33& 0.60 & 0.71 & 1.03 & 0.73 &  & 0.36 & 0.68 & 1.73 & 1.18 &  & 0.29 & 0.46 & 1.73 & 0.80 & - \\
		$\alpha_{2}$-NbSi$_2$As$_4$ &33& 0.62 & 0.63 & 1.17 & 0.74 &  & 0.39 & 0.74 & 2.11 & 1.56 &  & 0.27 & 0.43 & 2.11 & 0.91 & - \\
		$\alpha_{1}$-NbGe$_2$N$_4$ &33& 1.28 & 1.38 & 1.25 & 1.73 &  & 0.44 & 1.04 & 2.37 & 2.46 &  & 0.28 & 0.62 & 2.37 & 1.47 & 0.63 \\
		$\alpha_{2}$-NbGe$_2$P$_4$ &33& 0.63 & 0.72 & 1.06 & 0.76 &  & 0.37 & 0.71 & 1.87 & 1.33 &  & 0.28 & 0.48 & 1.87 & 0.90 & - \\
		$\alpha_{2}$-NbGe$_2$As$_4$ &33& 0.60 & 0.61 & 1.16 & 0.70 &  & 0.38 & 0.70 & 2.05 & 1.44 &  & 0.27 & 0.35 & 2.05 & 0.72 & - \\
		$\alpha_{1}$-TaSi$_2$N$_4$ &33& 1.12 & 1.57 & 0.87 & 1.37 &  & 0.35 & 1.12 & 1.65 & 1.85 &  & 0.30 & 0.71 & 1.65 & 1.17 & - \\
		$\alpha_{1}$-TaSi$_2$P$_4$ &33& 0.56 & 0.81 & 0.97 & 0.79 &  & 0.33 & 0.67 & 1.63 & 1.09 &  & 0.27 & 0.44 & 1.63 & 0.72 & - \\
		$\alpha_{2}$-TaSi$_2$As$_4$ &33& 0.55 & 0.67 & 1.10 & 0.74 &  & 0.37 & 0.73 & 1.91 & 1.39 &  & 0.24 & 0.33 & 1.91 & 0.63 & - \\
		$\alpha_{1}$-TaGe$_2$N$_4$ &33& 1.26 & 1.42 & 1.05 & 1.49 &  & 0.37 & 0.87 & 2.07 & 1.80 &  & 0.31 & 0.66 & 2.07 & 1.37 & 0.37 \\
		$\alpha_{2}$-TaGe$_2$P$_4$ &33& 0.57 & 0.82 & 0.97 & 0.80 &  & 0.33 & 0.67 & 1.70 & 1.14 &  & 0.27 & 0.44 & 1.70 & 0.75 & - \\
		$\alpha_{2}$-TaGe$_2$As$_4$ &33& 0.55 & 0.70 & 1.07 & 0.75 &  & 0.35 & 0.68 & 1.86 & 1.26 &  & 0.28 & 0.46 & 1.86 & 0.86 & - \\
			\rule{0pt}{4mm}%
		$\beta_{2}$-NbSi$_2$N$_4$ &33& 0.39 & 1.20 & 0.36 & 0.43 &  & 0.39 & 1.05 & 0.61 & 0.64 &  & 0.30 & 0.72 & 0.77 & 0.55 & - \\
		$\beta_{2}$-NbSi$_2$P$_4$ &33& 0.15 & 0.25 & 0.80 & 0.20 &  & 0.29 & 0.49 & 1.31 & 0.64 &  & 0.25 & 0.31 & 1.75 & 0.54 & - \\
		$\beta_{2}$-NbGe$_2$N$_4$ &33& 0.53 & 1.17 & 0.48 & 0.56 &  & 0.50 & 0.99 & 0.75 & 0.74 &  & 0.29 & 0.62 & 0.98 & 0.61 & - \\
		$\beta_{2}$-TaSi$_2$N$_4$ &33& 0.40 & 1.36 & 0.28 & 0.38 &  & 0.38 & 1.13 & 0.49 & 0.55 &  & 0.29 & 0.68 & 0.62 & 0.42 & - \\
		$\beta_{2}$-TaSi$_2$P$_4$ &33& 0.14 & 0.26 & 0.74 & 0.19 &  & 0.24 & 0.48 & 1.14 & 0.55 &  & 0.22 & 0.35 & 1.49 & 0.52 & - \\
		$\beta_{2}$-TaGe$_2$N$_4$ &33& 0.47 & 1.21 & 0.36 & 0.44 &  & 0.40 & 0.94 & 0.58 & 0.55 &  & 0.29 & 0.67 & 0.75 & 0.50 & - \\
			\rule{0pt}{4mm}%
            $\beta_{1}$-CrSi$_2$As$_4$ &34& - & - & - & - &  & 0.53 & 0.75 & 2.28 & 1.71 &  & 0.33 & 0.54 & 2.63 & 1.42 & 2.80 \\
		$\beta_{2}$-CrGe$_2$N$_4$  &34& - & - & - & - &  & 2.22 & 1.64 & 1.84 & 3.02 &  & 0.74 & 1.08 & 2.37 & 2.56 & 2.32 \\
		$\beta_{1}$-CrGe$_2$As$_4$ &34& - & - & - & - &  & 0.23 & 0.32 & 3.93 & 1.26 &  & 0.17 & 0.21 & 4.50 & 0.95 & 2.88 \\
			\rule{0pt}{4mm}%
		$\beta_{5}$-MnSb$_2$Se$_4$ &41& - & - & - & - &  & 0.56 & 0.33 & 6.50 & 2.15 &  & 0.59 & 0.60 & 7.90 & 4.74 & 3.38 \\
		$\beta_{5}$-MnSb$_2$Te$_4$ &41& - & - & - & - &  & 1.41 & 0.59 & 4.09 & 2.41 &  & 0.80 & 0.50 & 5.10 & 2.55 & 4.21 \\
		$\beta_{5}$-MnBi$_2$Se$_4$ &41& - & - & - & - &  & 1.21 & 0.56 & 6.30 & 3.53 &  & 1.00 & 0.77 & 7.87 & 6.06 & 3.52 \\
		$\beta_{5}$-MnBi$_2$Te$_4$ &41& - & - & - & - &  & 1.73 & 0.73 & 3.76 & 2.74 &  & 1.03 & 0.58 & 4.73 & 2.74 & 4.27
	\end{tabular}
\label{table3}
\end{ruledtabular}
\end{table*}

For the 34-valence-electron Cr-based compounds, we restrict our analysis to the three-orbital 
and five-orbital correlated subspaces, which more accurately reflect the multiorbital nature of 
the Cr $d$-states near the Fermi level. Within these subspaces, the degree of electronic correlation 
varies significantly with chemical composition. In particular, the nitrogen-based $\beta_2$-CrGe$_2$N$_4$ 
displays signatures of strong electronic correlation in the three-orbital model, with \( U/W \) = 2.2, consistent with 
a narrow bandwidth and reduced screening. This contrasts with the arsenic-based analogue, which 
exhibits broader bands and enhanced hybridization, leading to more delocalized Cr $d$-states and 
weaker correlations. The comparison between N- and As-based systems highlights the role of chemical
environment in tuning orbital localization and screening efficiency. These findings indicate that 
strong correlation effects are primarily confined to the N-based Cr compound, while the As-based
system lies in a more moderately correlated regime.

In the case of the 41-valence-electron Mn-based topological compounds, we likewise focus on the 
three- and five-orbital subspaces to capture the multiorbital character of the Mn-derived states 
near the Fermi level. Within these subspaces, the correlation strength \( U/W \) lies between 
approximately 0.6 and 1.7, with slightly reduced values in the five-orbital model due to enhanced 
hybridization with chalcogen and pnictogen \( p \)-orbitals. Although the overall correlation 
strength remains moderate, the combination of sizable on-site Coulomb repulsion $U$ and finite 
exchange \( J \) is sufficient to stabilize the experimentally observed ferromagnetic semiconducting 
phases in these compounds~\cite{Guo2024,Otrokov2019}.

Overall, the correlation strength trends across the 33-, 34-, and 41-valence-electron 
MA\(_2\)Z\(_4\) systems reveal a clear evolution governed by orbital localization, 
electron count, and screening efficiency. The 33-electron materials exhibit strong 
correlations in narrow half-filled bands, while the 34- and 41-electron systems show 
more moderate interactions depending on the degree of orbital delocalization. These 
insights provide a quantitative basis for analyzing magnetic instabilities, which we 
address next using the Stoner model.

While the correlation strength quantified by \( U/W \) provides valuable insight into the
electronic interactions in MA$_2$Z$_4$ compounds, it does not fully capture the tendency 
toward magnetic ordering, particularly in metallic systems with a finite density of states 
at the Fermi level. To complement the \( U/W \) analysis, we therefore turn to the Stoner 
model of itinerant magnetism, which offers a quantitative criterion for the onset of ferromagnetic 
instability. This framework is especially pertinent for the 33, 34, and 41 valence electron
compounds studied here, where magnetism arises from a single transition metal sublattice: V,
Nb, or Ta in the 33-valence-electron systems; Cr in the 34-valence-electron systems; and Mn 
in the 41-valence-electron materials. Our methodology builds on prior work that employed 
constrained RPA-derived Stoner parameters to analyze spin splitting and magnetic order in 
half-Heusler compounds with multiple magnetic sublattices~\cite{sasioglu2025itinerant}.

Magnetic instabilities in low-dimensional systems often emerge from a complex interplay 
between electron-electron interactions, density of states, and the symmetry of the relevant 
electronic orbitals. In the MA\(_2\)Z\(_4\) family, where magnetism originates from a single 
transition-metal sublattice, these instabilities are generally of itinerant character. For
example, our supercell DFT calculations for \(\alpha_1\)-VSi\(_2\)N\(_4\) reveal a substantial 
reduction in the V magnetic moment, by nearly 50\%—in an antiferromagnetic configuration compared to 
the ferromagnetic ground state. This pronounced sensitivity of the local moment to the 
magnetic ordering pattern is a hallmark of itinerant magnetism, in contrast to the localized 
magnetic moments of Mott systems. To quantitatively assess the tendency toward such 
itinerant magnetic ordering, we now turn to the Stoner model, which provides a well-defined 
criterion for the onset of ferromagnetism in metallic systems.

The Stoner criterion states that a system becomes unstable toward ferromagnetism when the product
\( I N(E_F) \) exceeds unity, where \( I \) is the Stoner parameter and \( N(E_F) \) is the density
of states at the Fermi level in the non-magnetic state. In this work, the Stoner parameter \( I \)
is obtained from the Slater parameterization of the cRPA Coulomb interaction matrix (see Methods 
section), using expressions tailored to each correlated subspace: \( I = \frac{1}{5}(U_s + 6J_s) \) 
for the full five-orbital \( d \)-manifold, \( I = \frac{1}{3}(U_s + 4J_s) \) for the three-orbital 
model, and \( I = U_s \) for the one-orbital case, where Hund’s coupling does not contribute. These 
formulas provide a consistent and physically grounded approach to estimating \( I \) from 
cRPA-derived interaction parameters across different orbital models and materials.

For the 33-valence-electron compounds, magnetic instabilities are most appropriately 
analyzed using the one-orbital model in all structural phases except the \(\beta_2\) phase. 
This choice is supported by the orbital-resolved band structures, which reveal that a single
orbital—predominantly \( d_{z^2} \)—dominates the electronic states near the Fermi level 
and contributes most significantly to the magnetic moment, as also reflected in the final 
column of Table~\ref{table3}. Within this one-orbital subspace, all V-based compounds 
satisfy the Stoner criterion \( I N(E_F) > 1 \) and consequently develop magnetic moments
close to \( 1~\mu_B \). In contrast, only a few Nb- and Ta-based compounds satisfy the 
criterion, and even in these cases, the resulting magnetic moments are significantly smaller, 
typically below \( 0.5~\mu_B \). These reduced magnetic moments are consistent with the smaller exchange 
splittings observed in spin-polarized DFT calculations (see Fig.~S5 in the Supplementary 
Information). A notable exception is \(\alpha_1\)-TaSi\(_2\)N\(_4\), which satisfies the 
Stoner criterion with \( I N(E_F) = 1.37 \) in the one-orbital model, yet remains nonmagnetic 
in spin-polarized DFT. This discrepancy underscores a limitation of the Stoner model when 
used with cRPA-derived interaction parameters. Specifically, the Stoner expressions employed
here are derived from the Hartree–Fock solution of the multiorbital Hubbard model~\cite{Stollhoff_1990} 
and neglect dynamic correlation effects, which are known to reduce the Stoner 
parameter by up to 40\%. Such correlations may shift the actual Stoner product below the 
critical threshold, thereby explaining the absence of magnetism in \(\alpha_1\)-TaSi\(_2\)N\(_4\).

For the \(\beta_2\)-phase structures of the 33-valence-electron compounds, the one-orbital model 
becomes inadequate due to the broader distribution of electronic states near the Fermi level. 
Instead, the three-orbital model provides a more accurate representation of the relevant correlated
subspace. Within this model, none of the Nb- and Ta-based \(\beta_2\)-phase compounds satisfy the 
Stoner criterion, consistent with their paramagnetic ground states observed in spin-polarized DFT 
calculations. In contrast, \(\beta_2\)-VGe\(_2\)N\(_4\) emerges as the only V-based compound in 
the \(\beta_2\) phase that fulfills the Stoner criterion in the three-orbital model, while failing 
to do so in the one-orbital case. This behavior highlights the spatially extended nature of the
magnetic orbitals and underscores the necessity of an enlarged correlated subspace for an accurate 
description of magnetic tendencies in this system.

Turning to the 34- and 41-valence-electron compounds, the three-orbital subspace provides the most 
appropriate description of magnetism, as the correlated states near the Fermi level involve multiple 
orbitals beyond the \( d_{z^2} \) component. For this reason, we do not report one-orbital results for 
both the Cr- and Mn-based compounds. As shown in Table~\ref{table3}, all Cr- and Mn-based systems satisfy 
the Stoner criterion \( I N(E_F) > 1 \) in the three-orbital model, indicating a strong tendency toward 
ferromagnetic ordering. The Mn-based layered topological compounds exhibit particularly large \( I N(E_F) \) 
values exceeding 2, along with magnetic moments above \( 3~\mu_B \), consistent with their well-established 
ferromagnetic semiconducting character. The three Cr-based compounds, \(\beta_2\)-CrGe\(_2\)N\(_4\),
\(\beta_1\)-CrSi\(_2\)As\(_4\), and \(\beta_1\)-CrGe\(_2\)As\(_4\), exhibit Stoner instabilities with 
\( I N(E_F) > 1 \) and develop sizable magnetic moments in spin-polarized DFT calculations. These findings 
confirm the reliability of the Stoner model in predicting itinerant ferromagnetism in Cr- and Mn-based 
MA\(_2\)Z\(_4\) systems and emphasize the role of multi-orbital correlations in stabilizing their magnetic 
ground states.

In summary, the Stoner model analysis across the MA\(_2\)Z\(_4\) family highlights the crucial 
role of both electron count and orbital character in governing magnetic instabilities. For 
33-valence-electron compounds, the magnetism is predominantly driven by a single localized orbital, 
whereas in the 34- and 41-electron systems, multi-orbital effects and enhanced density of states 
at the Fermi level lead to robust ferromagnetic tendencies. The agreement between Stoner criterion 
predictions and spin-polarized DFT results is generally strong, but deviations, such as in 
\(\alpha_1\)-TaSi\(_2\)N\(_4\), reveal the limitations of the 
Hartree-Fock-based approach when many-body correlation effects are significant. Despite these 
limitations, the combined \( U/W \) and Stoner analyses offer a comprehensive framework for 
understanding the correlation-driven magnetism in this diverse materials family, laying the 
foundation for subsequent exploration of their collective spin excitations.

\subsection*{Plasmon dispersion in cold metallic compounds}

\begin{figure*}[tp] 	
\centering
\includegraphics[width=1.0\linewidth]{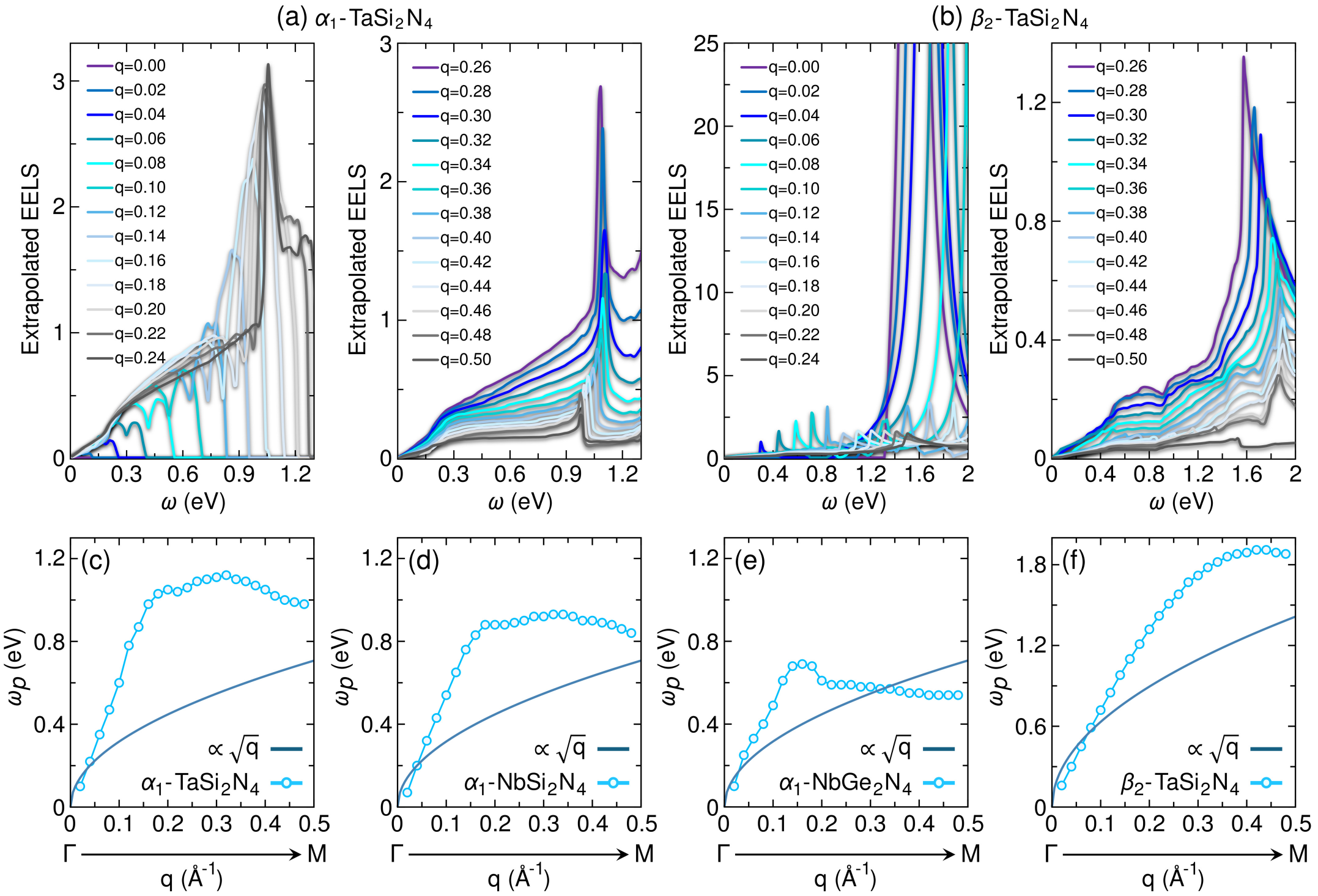}
\vspace{-0.5 cm}
\caption{Plasmonic properties of selected 33-valence-electron MA$_2$Z$_4$ cold metallic compounds. 
Panels (a) and (b) show the calculated electron energy loss spectra $L(\mathbf{q},\omega)$ for wavevectors 
along the $\Gamma$--M direction in (a) $\alpha_1$-TaSi$_2$N$_4$ and (b) $\beta_2$-TaSi$_2$N$_4$. 
Plasmon dispersion relations extracted from the peak positions in $L(\mathbf{q},\omega)$ are plotted in 
(c)--(f) for four representative materials: (c) $\alpha_1$-TaSi$_2$N$_4$, (d) $\alpha_1$-NbSi$_2$N$_4$, 
(e) $\alpha_1$-NbGe$_2$N$_4$, and (f) $\beta_2$-TaSi$_2$N$_4$. The results show pronounced deviations from 
conventional $\sqrt{q}$ plasmon dispersion, including linear and negative-slope regimes characteristic 
of cold-metal band structures.}
\label{fig:EELS}
\end{figure*}

Plasmons—collective oscillations of the electron density, play a central role in 
governing the optical and charge-response properties of low-dimensional systems. 
In two-dimensional materials, they exhibit strong spatial confinement, electrostatic 
tunability, and enhanced near-field intensities, making them promising candidates for 
applications in photodetectors, sensors, modulators, and energy-harvesting technologies~\cite{grigorenko2012graphene,low2017polaritons}. 
However, the practical realization of such plasmonic devices is often hindered by 
dissipative losses due to Landau damping, phonon coupling, or disorder-induced scattering. 
A key strategy to suppress these losses is to eliminate low-energy electron–hole pair excitations. 
This condition can be naturally fulfilled in systems with a narrow metallic band energetically 
isolated by internal and external gaps, as recently observed in a class of 2D cold metals.

Motivated by recent observations, particularly the emergence of low-loss plasmons in cold metallic transition-metal 
dichalcogenides (TMDs) such as TaS$_2$, TaSe$_2$, and NbSe$_2$~\cite{Wezel,gjerding2017band,da2020universal}, 
we turn our attention to structurally related MA$_2$Z$_4$ monolayers with 33 valence electrons. We focus 
on four representative compounds: $\alpha_1$-TaSi$_2$N$_4$, $\alpha_1$-NbSi$_2$N$_4$, $\alpha_1$-NbGe$_2$N$_4$, 
and $\beta_2$-TaSi$_2$N$_4$, which exhibit either pn-type or n-type cold metallic behavior. The Nb-based 
systems, while exhibiting small exchange splittings in their magnetic ground states (see Fig. S5 in the
Supplementary Information), are treated in  the non-magnetic configuration to isolate plasmonic features 
arising from their band topology. All four compounds feature a narrow $d_{z^2}$-derived band crossing the 
Fermi level; in the three $\alpha_1$-phase pn-type cold metals, this band is flanked by well-defined energy 
gaps above and below $E_F$, whereas in the $\beta_2$-phase TaSi$_2$N$_4$, only the lower gap is present, 
characteristic of an n-type cold metal. This band topology suppresses interband transitions and Landau damping,
making these systems ideal candidates for supporting low-loss, unconventional plasmon modes.

To evaluate the plasmonic response and characterize the collective charge excitations in these systems, 
we compute the loss function $L(\mathbf{q},\omega) = -\mathrm{Im}[1/\epsilon_m(\mathbf{q},\omega)]$, 
where $\epsilon_m(\mathbf{q},\omega) = 1/\epsilon_{00}^{-1}(\mathbf{q},\omega)$ denotes the macroscopic 
dielectric function. Plasmon modes appear as peaks in  $L(\mathbf{q},\omega)$, and their dispersion is 
extracted by tracking the peak positions as a function of in-plane wavevector $q = |\mathbf{q}|$. 
Figures~\ref{fig:EELS}(a) and \ref{fig:EELS}(b) show representative loss spectra, while 
panels~\ref{fig:EELS}(c)--\ref{fig:EELS}(f) display the resulting dispersion relations. While three-dimensional 
metals exhibit finite plasmon energies at $q=0$ and conventional 2D metals follow $\omega_p \propto \sqrt{q}$ behavior,
the MA$_2$Z$_4$ cold metals show markedly different trends.

In particular, the $\alpha_1$-phase compounds exhibit a distinct plasmon dispersion featuring a nearly linear slope 
at small $q$, followed by an extended flat region, and ultimately a negative slope at larger momenta. This non-monotonic 
behavior is strongly suppressed in the $\beta_2$ phase. The underlying mechanism can be traced to the pn-type cold metallic
band structure (see Fig. S4 in the Supplementary Information): a narrow $d_{z^2}$-derived band crossing the Fermi 
level is energetically isolated by internal and external
gaps, which limit the phase space for electron–hole pair excitations and suppress interband transitions near the plasmon
energy. As demonstrated in recent EELS studies of 2H-TMDs such as 1H-TaS$_2$, 1H-TaSe$_2$, and 1H-NbSe$_2$~\cite{Wezel}, 
the flattening and eventual downturn of the plasmon dispersion result from a progressive weakening of the screening response 
at finite momentum. In that context, the initial plasmon dispersion is positive due to effective low-energy screening; 
however, as $q$ increases, interband transitions are no longer available to efficiently screen the plasmon, leading first
to a plateau and then to a decrease in plasmon energy. This mechanism closely parallels the behavior observed in our
pn-type MA$_2$Z$_4$ compounds. The $\beta_2$-phase TaSi$_2$N$_4$, by contrast, lacks the upper gap and exhibits conventional
n-type behavior, resulting in a more monotonic dispersion. Our findings thus position MA$_2$Z$_4$ cold metals as promising 
platforms for low-loss 2D plasmonics, where dispersion characteristics can be engineered through band-structure 
design and phase control.

We note, however, that the plasmon dispersions presented here are derived from DFT-based band structures, which are known to 
underestimate band gaps relative to many-body approaches such as the GW method. In our recent GW study of 2D cold metallic 
TMDs and 33-valence-electron MA$_2$Z$_4$ compounds~\cite{Beida_2025}, we showed that both internal and external gaps are 
significantly enhanced compared to their
DFT counterparts. As the plasmon behavior in cold metals is strongly tied to the energetic separation between the narrow 
metallic band and surrounding states, larger GW gaps would further reduce interband screening channels and likely enhance 
the flattening or downturn of the dispersion. These considerations suggest that the unconventional plasmon features reported 
here may be even more pronounced when many-body corrections are taken into account.

While our analysis has focused on four representative compounds—$\alpha_1$-TaSi$_2$N$_4$, $\alpha_1$-NbSi$_2$N$_4$, 
$\alpha_1$-NbGe$_2$N$_4$, and $\beta_2$-TaSi$_2$N$_4$—several other 33-valence-electron MA$_2$Z$_4$ monolayers listed 
in Table~\ref{table2} exhibit similar electronic features. Notable examples include $\alpha_1$-TaGe$_2$N$_4$, 
$\beta_2$-VGe$_2$N$_4$, and $\alpha_2$-VGe$_2$P$_4$, all of which possess narrow $d_{z^2}$ bands crossing $E_F$ 
and display pn-type cold metallic character. Their similarly restricted interband phase space suggests they may also 
support low-loss plasmon modes with linear and negative dispersion regimes. A broader theoretical and experimental 
exploration of these systems could open new avenues for designing tunable 2D plasmonic materials based on pn-type 
cold metals.

Overall, our results provide a comprehensive mapping of effective Coulomb interactions and their implications 
across semiconducting, metallic, cold-metallic, magnetic, and topological MA$_2$Z$_4$ monolayers. Using cRPA, 
we computed material-specific interaction parameters and revealed unconventional screening behavior in 
both semiconductors and cold-metallic systems. Our analysis points to the emergence of strong-coupling 
tendencies in several 33- and 34-valence-electron compounds, including signatures of proximity to Mott 
or charge-density-wave instabilities in narrow-band phases. In addition, we identified magnetic instabilities 
in V-, Nb-, Cr-, and Mn-based monolayers through a subspace-resolved Stoner criterion. Finally, our 
exploration of a selected set of cold-metallic compounds uncovered anomalous plasmon dispersions 
characterized by nearly non-dispersive low-energy modes. These findings position MA$_2$Z$_4$ monolayers
as a fertile platform for investigating screening, correlation-driven instabilities, and low-energy 
collective behavior in two dimensions.

\section*{Methods}

\subsection*{Computational details}

Our first-principles calculations are based on DFT within the full-potential
linearized augmented plane wave (FLAPW) framework, as implemented in the \textsc{FLEUR} code~\cite{FLEUR}. 
The exchange-correlation functional is treated using the generalized gradient approximation (GGA) in the
Perdew–Burke–Ernzerhof (PBE) parameterization~\cite{Perdew}. We perform both non-spin-polarized and
spin-polarized DFT calculations where appropriate: non-magnetic calculations are used for all cRPA 
evaluations, while spin-polarized calculations are carried out only for compounds investigated using the 
Stoner model of magnetic instability. All calculations are based on experimentally reported or previously
published lattice structures and atomic positions, without any additional structural 
relaxation~\cite{Wang,Zhu2021,Wimmer2021}.

For the self-consistent DFT calculations, we use a $16 \times 16 \times 1$ $\mathbf{k}$-point mesh in the 
2D Brillouin zone and a plane-wave cutoff of $G_{\mathrm{max}} = 4.5$~bohr$^{-1}$ for the interstitial
wavefunctions. Wannier projectors are constructed for the transition metal $d$ orbitals using the \textsc{Wannier90} 
package~\cite{Marzari}, interfaced with \textsc{FLEUR} code. The number and character of the Wannier functions are 
adapted to the orbital content near the Fermi level in each system, with particular attention to one-, three-, 
and five-orbital correlated subspaces.

The outputs of the DFT calculations serve as input for the \textsc{SPEX} code~\cite{friedrich2010efficient,schindlmayr2010first,SPEX2}, which we use
to evaluate the partially screened and fully screened Coulomb interaction matrices within the constrained random
phase approximation (cRPA) and the full RPA. For these calculations, we employ a $12 \times 12 \times 1$
$\mathbf{k}$-point grid and include sufficient unoccupied states to ensure convergence of the polarization function
and dielectric screening.

\subsection*{Overview of the cRPA formalism}

The constrained random phase approximation (cRPA) provides an \textit{ab initio} framework for computing the effective Coulomb interaction among localized electrons, while systematically excluding screening processes arising within the correlated subspace~\cite{csacsiouglu2011effective,aryasetiawan2004frequency,Nomura2012,vaugier2012hubbard}. This method yields frequency-dependent matrix elements of the partially screened interaction \( U(\omega) \), including on-site, interorbital, exchange, and long-range components.

In the cRPA formalism, the total polarization function \( P \) is decomposed into two parts: \( P = P_d + P_r \), where \( P_d \) includes only transitions within the correlated subspace, and \( P_r \) accounts for the remaining screening channels. The partially screened Coulomb interaction is then obtained as:
\begin{equation}
U(\omega) = \left[ 1 - v P_r(\omega) \right]^{-1} v,
\end{equation}
where \( v \) is the bare Coulomb interaction. The fully screened interaction \( \tilde{U}(\omega) \) is subsequently obtained by applying the residual screening from the correlated subspace:
\begin{equation}
\tilde{U}(\omega) = \left[ 1 - U(\omega) P_d(\omega) \right]^{-1} U(\omega).
\end{equation}

The static matrix elements \( U \) are evaluated in the basis of maximally localized Wannier functions \( w_{i n}(\mathbf{r}) \) as:
\begin{align}
U_{i n_1, j n_3; i n_2, j n_4} &= 
\int d\mathbf{r} \int d\mathbf{r}'\,
w^{*}_{i n_1}(\mathbf{r})\, w^{*}_{j n_3}(\mathbf{r}') \nonumber \\
&\quad \times U(\mathbf{r}, \mathbf{r}', \omega = 0)\,
w_{j n_4}(\mathbf{r}')\, w_{i n_2}(\mathbf{r}).
\label{eq:cRPA_matrix}
\end{align}

In this work, we consider three distinct correlated subspaces: a one-orbital model (typically dominated by a localized \( d_{z^2} \)-like orbital), a three-orbital model, and the full five-orbital \( d \)-manifold. To characterize the effective interaction strength in these subspaces, we extract average Coulomb parameters using two standard parameterizations. For most of the analysis in this work, including the determination of correlation strength via \( U/W \), we adopt the Hubbard–Kanamori parametrization, which provides intraorbital (\( U \)), interorbital (\( U' \)), and  exchange (\( J \)) interactions defined by:
\begin{align}
U &= \frac{1}{L} \sum_{m} U_{m m; m m}, \label{eq:Uavg} \\
U' &= \frac{1}{L(L-1)} \sum_{m \ne n} U_{m n; m n}, \label{eq:Uprime} \\
J &= \frac{1}{L(L-1)} \sum_{m \ne n} U_{m n; n m}, \label{eq:Javg}
\end{align}
where \( L = 1 \), 3, or 5 denotes the number of correlated orbitals in the subspace. These expressions are used to extract orbital-averaged interactions consistent with multiorbital Hubbard or Kanamori-type Hamiltonians.
In addition to the partially screened parameters discussed above, one can also define the corresponding fully screened (\( \tilde{U} \), \( \tilde{U}' \), \( \tilde{J} \)) and bare (\( V \), \( V' \), \( J_V \)) Coulomb parameters using the same averaging scheme. The fully screened interaction \( \tilde{U} \) includes all screening channels, while the bare interaction \( V \) is calculated from the unscreened Coulomb kernel and reflects the spatial extent and localization of the Wannier orbitals. Together, these different levels of screening provide a comprehensive picture of the interaction landscape and its material dependence.

In addition, to calculate the Stoner parameter \( I \), we use the Slater parametrization of the Coulomb matrix. This orbital-averaged form simplifies the interaction to number-operator terms and is given by:
\begin{align}
U_S &= \frac{1}{L^2} \sum_{m, n} U_{m n; m n}, \label{eq:US} \\
J_S &= U_S - \frac{1}{L(L-1)} \sum_{m \ne n} 
\left( U_{m n; m n} - U_{m n; n m} \right). \label{eq:JS}
\end{align}
This parametrization ensures consistency with mean-field models of itinerant magnetism and enables a subspace-resolved evaluation of magnetic instabilities.

Together, the Hubbard–Kanamori and Slater parameterizations provide a consistent framework for quantifying electron correlation strength, analyzing orbital-dependent interactions, and identifying the onset of magnetic tendencies across the chemically and electronically diverse MA\(_2\)Z\(_4\) monolayers.

Calculating Coulomb interaction parameters for the various structural phases of MA\(_2\)Z\(_4\) monolayers is computationally feasible. However, in charge-density-wave (CDW) phases with star-of-David (SOD) reconstructions, the unit cell expands significantly, comprising 91 atoms, including 13 transition metal (TM) sites. Direct cRPA calculations for these large supercells are prohibitively expensive. To estimate the effective on-site interaction in such systems, we average the long-range interaction over all TM–TM site pairs within a single star cluster of the corresponding undistorted \( \beta \) structure:
\begin{equation}
\label{U_eff_SOD}
U^{\mathrm{eff}} = \frac{1}{13^2} \sum_{\mathbf{R}, \mathbf{R}^\prime} U_{\mathbf{R} - \mathbf{R}^\prime},
\end{equation}
where \( \mathbf{R} \) and \( \mathbf{R}^\prime \) label the positions of TM atoms within the star. This average defines the effective on-site interaction \( U_{00}^{\mathrm{eff}} \) for that star~\cite{Kim-2023,kamil_2018}. 

To estimate the effective intersite interaction \( U_{01}^{\mathrm{eff}} \) between neighboring stars in the reconstructed CDW lattice, one instead averages over all TM site pairs where \( \mathbf{R} \) belongs to star A and \( \mathbf{R}^\prime \) belongs to the nearest-neighboring star B (see Fig.~\ref{fig:struc}(f) for a schematic of the SOD geometry). This approximation captures the key interaction scales relevant for describing electronic correlations in the SOD phase.

\section*{Data Availability}
All data supporting the findings of this study are available from the authors upon reasonable request.

\section*{Competing Interests}
The authors declare no competing interests.

\begin{acknowledgments}
This work was supported by the Iran National Science Foundation (INSF) under Project No. 4044100,  
the Collaborative Research Center CRC/TRR 227 of the Deutsche Forschungsgemeinschaft (DFG),  
and the European Union (EFRE) via Grant No. ZS/2016/06/79307.
\end{acknowledgments}


%

\end{document}